\documentclass[longauth,twocolumns]{aa}  
\usepackage{graphicx}
\usepackage{txfonts}
\usepackage{natbib,twoopt}
\usepackage[breaklinks=true]{hyperref} 
\bibpunct{(}{)}{;}{a}{}{,} 
\makeatletter
\newcommandtwoopt{\citeads}[3][][]{\href{http://adsabs.harvard.edu/abs/#3}%
{\def\hyper@linkstart##1##2{}%
\let\hyper@linkend\@empty\citealp[#1][#2]{#3}}}
\newcommandtwoopt{\citepads}[3][][]{\href{http://adsabs.harvard.edu/abs/#3}%
{\def\hyper@linkstart##1##2{}%
\let\hyper@linkend\@empty\citep[#1][#2]{#3}}}
\newcommandtwoopt{\citetads}[3][][]{\href{http://adsabs.harvard.edu/abs/#3}%
{\def\hyper@linkstart##1##2{}%
\let\hyper@linkend\@empty\citet[#1][#2]{#3}}}
\newcommandtwoopt{\citeyearads}[3][][]%
{\href{http://adsabs.harvard.edu/abs/#3}
{\def\hyper@linkstart##1##2{}%
\let\hyper@linkend\@empty\citeyear[#1][#2]{#3}}}
\makeatother

\newcommand{\kms}{\ensuremath{{\rm km\,s^{-1}}}}                     

\begin{document} 

   \title{VLTI-AMBER  velocity-resolved aperture-synthesis imaging of $\eta$~Carinae with a spectral resolution of 12\,000.}
   \subtitle{Studies  of the primary star wind and innermost wind-wind collision zone\thanks{Based on observations collected at the European Organisation for Astronomical Research in the Southern Hemisphere under ESO program 092.D-0289(A).   }    
   \thanks{The images in Fig.~\ref{f} are available in electronic form
at the CDS via anonymous ftp to cdsarc.u-strasbg.fr   (130.79.128.5)
or via http://cdsweb.u-strasbg.fr/cgi-bin/qcat?J/A+A/}
    }

   \author{G.~Weigelt\inst{1} 
\and K.-H.~Hofmann\inst{1}
\and D.~Schertl\inst{1}
\and N.~Clementel\inst{2}
\and M.F.~Corcoran\inst{3,4}
\and A.~Damineli\inst{5}
\and W.-J.~de~Wit\inst{6}
\and R.~Grellmann\inst{7}
\and J.~Groh\inst{8}
\and S.~Guieu\inst{6}
\and T.~Gull\inst{9}
\and M.~Heininger\inst{1}
 \and D.J.~Hillier\inst{10}
 \and C.A.~Hummel\inst{11}
\and S.~Kraus\inst{12}
\and T.~Madura\inst{9}
\and A.~Mehner\inst{6}
\and A.~M\'erand\inst{6}
\and F.~Millour\inst{13} 
\and A.F.J.~Moffat\inst{14}
\and K.~Ohnaka\inst{15}
\and F.~Patru\inst{16}
\and R.G.~Petrov\inst{13}
\and S.~Rengaswamy\inst{17}
\and N.D.~Richardson\inst{18}
\and T.~Rivinius\inst{6}
\and M.~Sch\"oller\inst{11}
\and M.~Teodoro\inst{9}
\and M.~Wittkowski\inst{11}                   }
       
\institute{Max Planck Institute for Radio Astronomy, Auf dem H\"ugel 69, 53121 Bonn, Germany  \\     \email{weigelt@mpifr.de}
\and South African Astronomical Observatory, PO box 9, 7935, Observatory, South Africa 
\and CRESST and X-ray Astrophysics Laboratory, Goddard Space Flight Center, Greenbelt, MD 20771, USA
\and Universities Space Research Association, 10211 Wincopin Circle, Suite 500, Columbia, MD 21044, USA
\and Instituto de Astronomia, Geof\'isica e Ci\^encias Atmosf\'ericas, Universidade de S\~ao Paulo, Rua do Mat\~ao 1226, Cidade Universit\'aria, S\~ao Paulo 05508-900, Brazil 
\and European Southern Observatory, Casilla 19001, Santiago 19,  Chile
\and I. Physikalisches Institut, Universit\"at zu K\"oln, Z\"ulpicher Strasse 77, 50937 K\"oln, Germany
\and School of Physics, Trinity College Dublin, The University of Dublin, Dublin 2, Ireland
 \and Code 667, Astrophysics Science Division, Goddard Space Flight Center, Greenbelt, MD 20771 USA 
\and Department of Physics and Astronomy \& Pittsburgh Particle Physics, Astrophysics,  and Cosmology Center (PITT PACC),  University of Pittsburgh, 3941 O'Hara Street, Pittsburgh, PA 15260, USA
\and European Southern Observatory, Karl Schwarzschild Strasse 2,  85748 Garching, Germany
\and University of Exeter, Astrophysics Group, Stocker Road, Exeter, EX4 4QL, UK
\and Laboratoire Lagrange, UMR7293, Universit\'e de Nice Sophia-Antipolis, CNRS, Observatoire de la C\^ote d'Azur, 06300 Nice, France
\and D\'epartement de physique and Centre de Recherche en Astrophysique du Qu\'ebec (CRAQ), Universit\'e de Montr\'eal, CP 6128 Succ. A., Centre-Ville, Montr\'eal, Qu\'ebec H3C 3J7, Canada
\and Universidad Cat\'olica del Norte, Instituto de Astronom\'ia, Avenida Angamos 0610, Antofagasta, Chile
\and  Osservatorio Astrofisico di Arcetri, 5 Largo Enrico Fermi, 50125, Firenze, Italia  
\and Indian Institute of Astrophysics, Koramangala, Bengaluru 560034, India  
\and Ritter Observatory, Department of Physics and Astronomy, The University of Toledo, Toledo, OH 43606-3390, USA
                          }

 \date{Received September 15, 1996; accepted March 16, 1997}

 
 \titlerunning{AMBER/VLTI observations of $\eta$\,Carinae}

\authorrunning{G.~Weigelt et al.\ }
 
  \abstract
   {The mass loss from massive stars is not understood well.  \object{$\eta$~Carinae} is a unique object for studying the massive stellar wind during the Luminous Blue Variable phase. 
   It is also an eccentric binary with a period of 5.54~yr. The nature of both stars is uncertain, although we know from X-ray studies that there is a wind-wind collision whose properties change with orbital phase.  }
   {We want to investigate the structure and kinematics of $\eta$~Car's primary star wind and wind-wind collision zone with a high spatial resolution of $\sim 6$~mas ($\sim14$~au) and high spectral resolution of $R = 12\,000$.       }  
   {Observations of \object{$\eta$~Car} were carried out with the ESO Very Large Telescope Interferometer (VLTI) and the AMBER instrument between approximately five and seven months before the August 2014 periastron passage. Velocity-resolved aperture-synthesis images were reconstructed from the spectrally dispersed interferograms. Interferometric studies can provide information on the binary orbit, the primary wind, and the wind collision.}
   {We present velocity-resolved aperture-synthesis images reconstructed in more than 100 different spectral channels distributed across the Br$\gamma$~2.166~$\mu$m  emission  line. The intensity distribution of the images strongly depends on wavelength.  At wavelengths corresponding to radial velocities of approximately $-$140 to $-376~\kms$  measured relative to line center, the intensity distribution has a fan-shaped structure. At the  velocity of  $-277~\kms$, the position angle of the  symmetry axis of the fan is $\sim 126\degr$.    The fan-shaped structure extends approximately 8.0~mas ($\sim18.8$~au) to the southeast  and 5.8~mas ($\sim13.6$~au) to the northwest,  measured  along the symmetry axis at the 16\% intensity contour.    The shape of the  intensity distributions  suggests that the obtained images are the first direct images of the innermost wind-wind collision zone.  Therefore, the observations provide  velocity-dependent image structures that can be used to test three-dimensional hydrodynamical, radiative transfer  models of the massive interacting winds of $\eta$~Car.     
      }    
   {}

  \keywords{Stars: winds, outflows --   stars: individual:  $\eta$ Carinae -- stars: massive -- stars: mass-loss -- binaries: general -- techniques: interferometric}
  \maketitle
      

\section{Introduction}    \label{i} 

Studies of the mass-loss process are of crucial importance for improving our understanding of stellar evolution. Infrared long-baseline interferometry provides us with a unique opportunity to study the mass loss from the  colliding wind binary  $\eta$~Car.  The primary star of the system  is an extremely luminous and massive ($M \sim 100~M_{\odot}$) unstable Luminous Blue Variable star  (LBV) with a high mass-loss rate of $\sim 10^{-3} M_\odot {\rm yr}^{-1}$ 
\citepads{1997ARA&A..35....1D,
2001AJ....121.1569D, 
2001ApJ...553..837H,
2006ApJ...642..1098H,
2012ApJ...759L...2G}.  
  A decrease in the observed strength of H$\alpha$ and \ion{Fe}{ii}  emission lines, and a change in the strength of \ion{N}{ii}   lines, provide indications that the primary wind density has decreased, either due to a global reduction in mass loss or to latitudinal changes    \citepads[e.g.,][]{
2015ApJ...801L..15D,
2012ApJ...751...73M,
2010ApJ...717L..22M}. 
 However,  there is contradictory evidence from analysis of the X-ray light curves  \citepads[e.g.,][]{2013MNRAS.436.3820M}  and the near constancy of \ion{He}{ii}\,0.4686\,$\mu$m \citepads[e.g.,][]{2016ApJ...819..131T}, which indicate much smaller changes in the mass-loss rate.   
 
 The distance of $\eta$~Car is $\sim 2.35$~kpc     \citepads{1993PASAu..10..338A,
1997ARA&A..35....1D,
2006ApJ...644.1151S}. 
  It is surrounded by the spectacular bipolar Homunculus nebula. The inclination of the polar axis of the Homunculus nebula with the line-of-sight is $\sim 41\degr$   \citepads{2001AJ....121.1569D, 
2006ApJ...644.1151S},  
 and the  position angle (PA)  of the projected Homunculus axis is $\sim 132\degr$  \citepads{1997ARA&A..35....1D,
2001AJ....121.1569D,
2006ApJ...644.1151S}.   

Using ESO's Very Large Telescope Interferometer (VLTI), the diameter of $\eta$~Car's  wind region was measured to be about 5~mas (50\% encircled-intensity diameter measured in the field-of-view, FOV, of 70~mas) in the $K$-band continuum \citepads{2003A&A...410L..37V,2007A&A...464.1045K,2007A&A...464...87W}. The measured visibilities are in good agreement with the predictions from the detailed spectroscopic model by \citetads{1998ApJ...496..407H}  and   \citetads{2001ApJ...553..837H,2006ApJ...642..1098H}. 
A good agreement between the Hillier model and interferometric observations of the LBV  P~Cyg was  reported by  \citetads{2013ApJ...769..118R}. 

First  spectro-interferometric VLTI-AMBER observations of $\eta$~Car \citepads{2007A&A...464...87W} with medium and high spectral resolution in the \ion{He}{I}~2.059~$\mu$m and the Br$\gamma~$2.166~$\mu$m emission lines allowed us  to study the spatial structure of $\eta$~Car's  wind as a function of wavelength  with a high spatial resolution of $\sim$~5~mas  and  spectral resolutions of 1500 and 12\,000. The measured line visibilities agree with predictions of the radiative transfer model of \citetads{2001ApJ...553..837H}. We derived 50\%  encircled-intensity diameters of 4.2~mas (9.9~au), 6.5~mas (15.3~au), and 9.6~mas  (22.6~au) in the 2.17~$\mu$m continuum, the \ion{He}{i}~2.059~$\mu$m, and the Br$\gamma$~2.166~$\mu$m emission lines, respectively (for comparison, the radius $R_{\ast}$ of the primary star of $\eta$~Car is  $\sim$~100~$R_{\sun}  \sim$ 0.47~au $\sim$ 0.20~mas; however, $R_{\ast}$ is not known well; \citeads{2001ApJ...553..837H}). 

\citetads{2003ApJ...586..432S} studied the stellar light reflected by the Homunculus nebula and found that the velocity of the primary stellar wind is higher near the south pole than at the latitude corresponding to our line-of-sight.  Other studies suggest that the spectroscopic and interferometric observations can be explained by the wind-wind collision zone  \citepads{
2012ApJ...759L...2G,
2012ApJ...751...73M}. 

Studies of the binary or binary wind-wind collision zone were reported by many authors \citepads[e.g.,][]{1996ApJ...460L..49D,
1997NewA....2..107D,
1998A&AS..133..299D,
2000ApJ...528L.101D,
2008MNRAS.384.1649D,
1997NewA....2..387D,
2005AJ....129..900D,
2015ApJ...801L..15D,
2000ApJ...529L..99S,
2004ApJ...605..405S,
2010MNRAS.402..145S,
2002A&A...383..636P,
1997Natur.390..587C,
2001ApJ...547.1034C,
2005AJ....129.2018C,
1999ApJ...524..983I,
2003ApJ...597..513S,
2007ApJ...661..482S,
2005AJ....129.1694W,
2006ApJS..163..173G,
2009MNRAS.396.1308G,
2011ApJ...743L...3G,
2016arXiv160806193G, 
2007ApJ...660..669N,
2007A&A...464...87W,
2007MNRAS.378.1609K,
2009NewA...14...11K,
2009MNRAS.397.1426K,
2009MNRAS.394.1758P,
2010A&A...517A...9G,
2012ApJ...759L...2G,
2012MNRAS.423.1623G,
2010RMxAC..38...52M,
2012MNRAS.420.2064M,
2013MNRAS.436.3820M,
2010ApJ...710..729M,
2011ApJ...740...80M,
2012ApJ...751...73M,
2015A&A...578A.122M,
2010AJ....139.1534R,
2013ApJ...773L..16T,
2016ApJ...819..131T,
2015MNRAS.450.1388C,
2015MNRAS.447.2445C,
2016ApJ...817...23H}.  
The X-rays are believed to arise from a  collision between the winds from the LBV and a hotter O or WR-type star. The CHANDRA X-ray spectrum can be explained \citepads{2002A&A...383..636P}  by the collision of the primary  star wind  ($\dot{M} = 2.5 \times10^{-4} M_\odot {\rm yr^{-1}}$, terminal velocity $\sim 500~\kms$) and the wind of a hot companion ($\dot{M} = 10^{-5} M_\odot {\rm yr^{-1}}$, terminal velocity $\sim 3000~\kms$; probably an extreme Of or a WR star perhaps similar to the three luminous WNLh stars seen elsewhere in the Carina Nebula).  

The wind collision leads to a wind-wind collision cavity  
\citepads[e.g.,][]{2002A&A...383..636P,
2008AJ....135.1249H, 
2008MNRAS.388L..39O,
2009MNRAS.394.1758P, 
2011ApJ...726..105P,
2009MNRAS.396.1308G,
2011ApJ...743L...3G,
2010RMxAC..38...52M,
2012MNRAS.420.2064M,
2013MNRAS.436.3820M,
2010A&A...517A...9G,
2012MNRAS.423.1623G,
2012ApJ...759L...2G}. 
  Radiative transfer models of the wind-wind collision zone were reported by \citetads{2015MNRAS.450.1388C,2015MNRAS.447.2445C}.  Three-dimensional (3D) smoothed particle hydrodynamic simulations were used to determine the orientation of the binary's orbit (semimajor axis length $\sim$~15.4~au) in space and on  the sky  \citepads[e.g.,][]{2008MNRAS.388L..39O,
2012MNRAS.420.2064M}. 
\citetads{2012MNRAS.420.2064M} derived an orbit inclination  of 130 to $145\degr$ and a PA of the orbital axis projected on the sky of  302 to $327\degr$.  Therefore, the orbital axis of the binary is closely aligned with the Homunculus system axis. The argument of periapsis places the secondary star and wind-wind collision zone on the observer's side of the system when the companion star is at apastron. 

Detailed HST/STIS imaging of $\eta$~Car  with a resolution of about 0.1$\arcsec$ allowed  
\citetads{2009MNRAS.396.1308G,2011ApJ...743L...3G}, 
\citetads{2010ApJ...710..729M},  
\citetads{2013ApJ...773L..16T}, 
and  \citetads{2016arXiv160806193G} 
to image the circumstellar environment of  $\eta$~Car  across emission lines and at several orbital phases. The discovered structures are the result of the wind-wind collision over the past $\sim$~20~yr. These structures are called fossil wind structures, because they were created over several orbital cycles in the past.

Various observations suggest  that disk-like or toroidal material  in our line-of-sight to the central object  partially obscures the central star \citepads[e.g.,][]{1992A&A...262..153H,1995RMxAC...2...11W,1995AJ....109.1784D,
1997AJ....113..335D,
2001ApJ...553..837H,
2002ApJ...567L..77S,
2004ApJ...605..405S,
2010ApJ...710..729M,
2013MNRAS.436.3820M}. 
 To study this occulting material,   \citetads{1996A&A...306L..17F}  performed speckle masking imaging polarimetry (in H$\alpha$  plus continuum).  The polarized flux image (Fig. 2b in the polarimetry paper) shows a  bar with a length of $\sim 0.4\arcsec$ along a PA of $\sim$ 45 and $225\degr$, which was interpreted as an obscuring equatorial disk. Interestingly, this polarized 45$\degr$ bar has approximately the same length and PA as the fossil wind bar discovered by  \citetads{2011ApJ...743L...3G,2016arXiv160806193G}.


\begin{table}                                                     
\small
\caption{Summary of observations}              
\label{list}                                                       
\centering                                              
\begin{tabular}{lrll}                                
\hline\hline                                             
Date      &N$_{\rm{obs}}$ \tablefootmark{a}     &N$_{\rm{vis}}$ \tablefootmark{b}  &Cal \\         
\hline                                                                                  
2013 Dec 31  &1     &2  &HD92305  \\    
2014 Jan 04       &1      &3   &HD28749  \\
2014 Jan 19       &2      &1   &HD18660 \\
2014 Jan 20       &9      &5   &HD23319 \\
2014 Jan 21       &10      &0   & - \\
2014 Feb 03       &1      &0  & - \\
2014 Feb 06       &1      &0  & - \\
2014 Feb 07       &1      &0  & - \\
2014 Feb 08       &1      &0  & - \\
2014 Feb 09       &1      &0  & - \\
2014 Feb 13       &1      &0  & - \\
2014 Feb 14       &1      &0  & - \\
2014 Feb 21       &2     &0  & - \\
2014 Feb 22       &3      &0  & - \\
2014 Feb 23       &1      &0  & - \\
2014 Feb 24       &1     &0  & - \\
2014 Feb 25      &1     &0  & - \\
2014 Feb 26       &2     &0  & - \\
2014 Feb 27       &3     &0  & - \\ 2014 Feb 28       &2     &0  & - \\ 2014 Mar 04       &2     &0  & - \\
2014 Mar 05&3     &3  & HD46933 \\              \hline    \end{tabular}  
\tablefoot{
\tablefootmark{a} {Number of observations at different hour angles per night. }
\tablefootmark{b}{Number of visibilities calibrated with an interferometric calibrator star.}      }   
\end{table}  

 
\begin{figure}   \centering \includegraphics[width=78mm,angle=0]{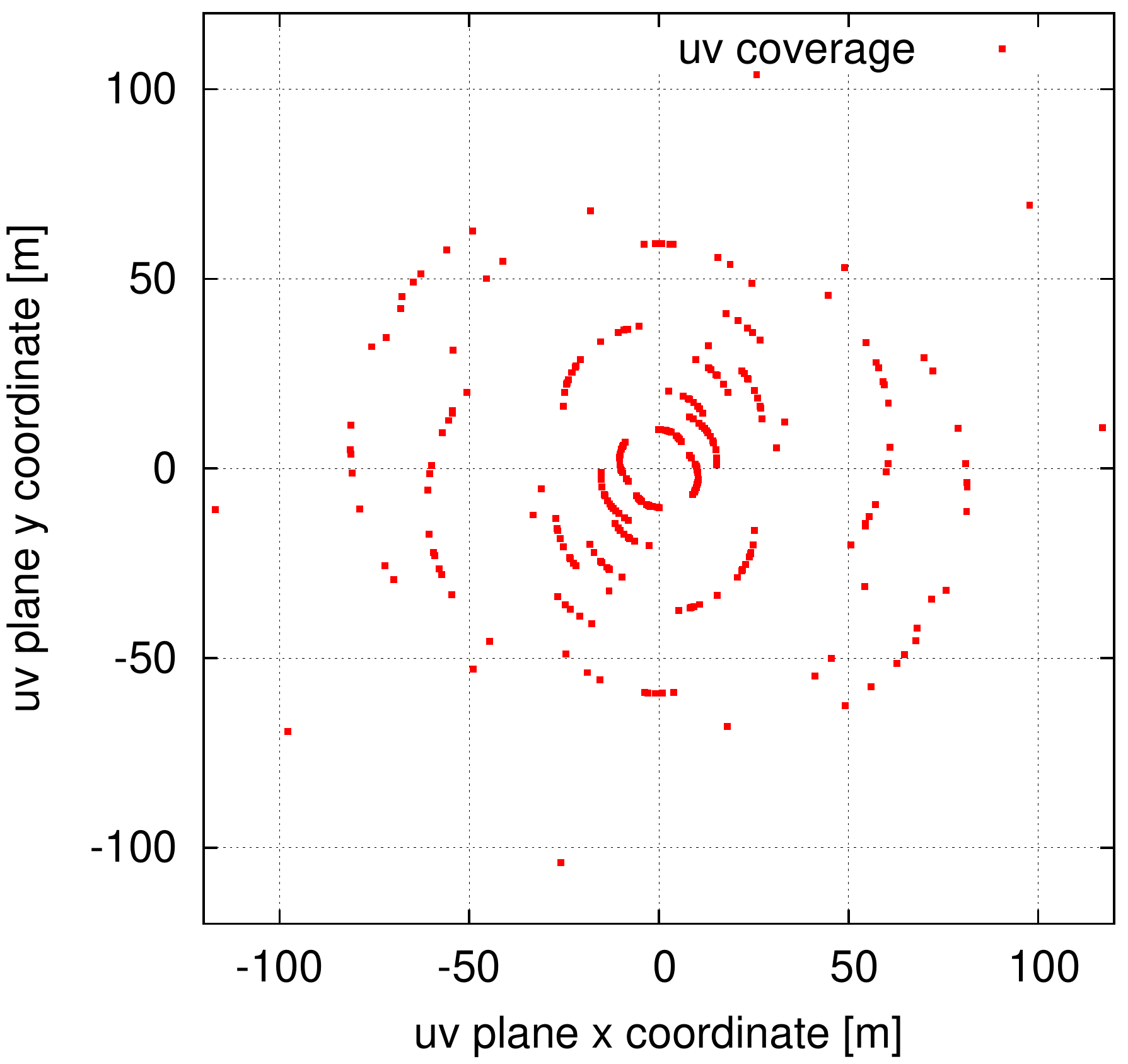}   \caption{$uv$ coverage of all VLTI/AMBER  observations listed in Table~\ref{list}. }   \label{uv} \end{figure}   


\begin{figure} \resizebox{\hsize}{!}{\includegraphics{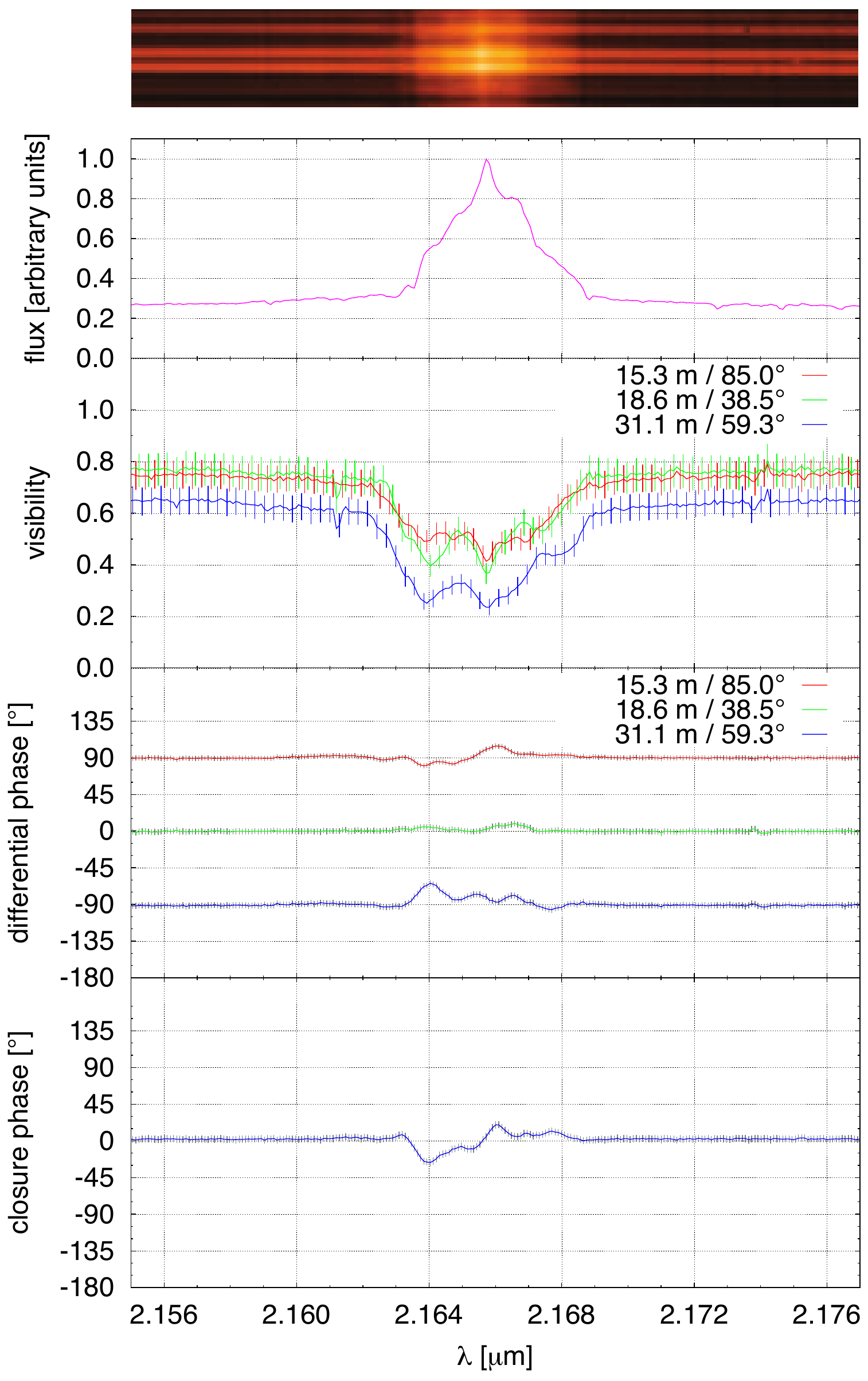}}
\caption{Illustrative example of an AMBER HR-mode ($R$\,=\,12\,000) observation of $\eta$~Car (see Sect.~\ref{o}).  The figure shows from top to bottom:  example interferogram,  spectrum of the  Br$\gamma$ region (heliocentric wavelengths; the telluric lines are not removed),  wavelength dependence of the visibilities at three baselines (PAs and projected baseline lengths are indicated in the panels),  wavelength-differential phases (the curve of the shortest baseline is shifted up by 90$\degr$ and the longest down by 90$\degr$),  and closure phases.  }      \label{obs1} \end{figure}


\begin{figure}  \resizebox{\hsize}{!}{\includegraphics{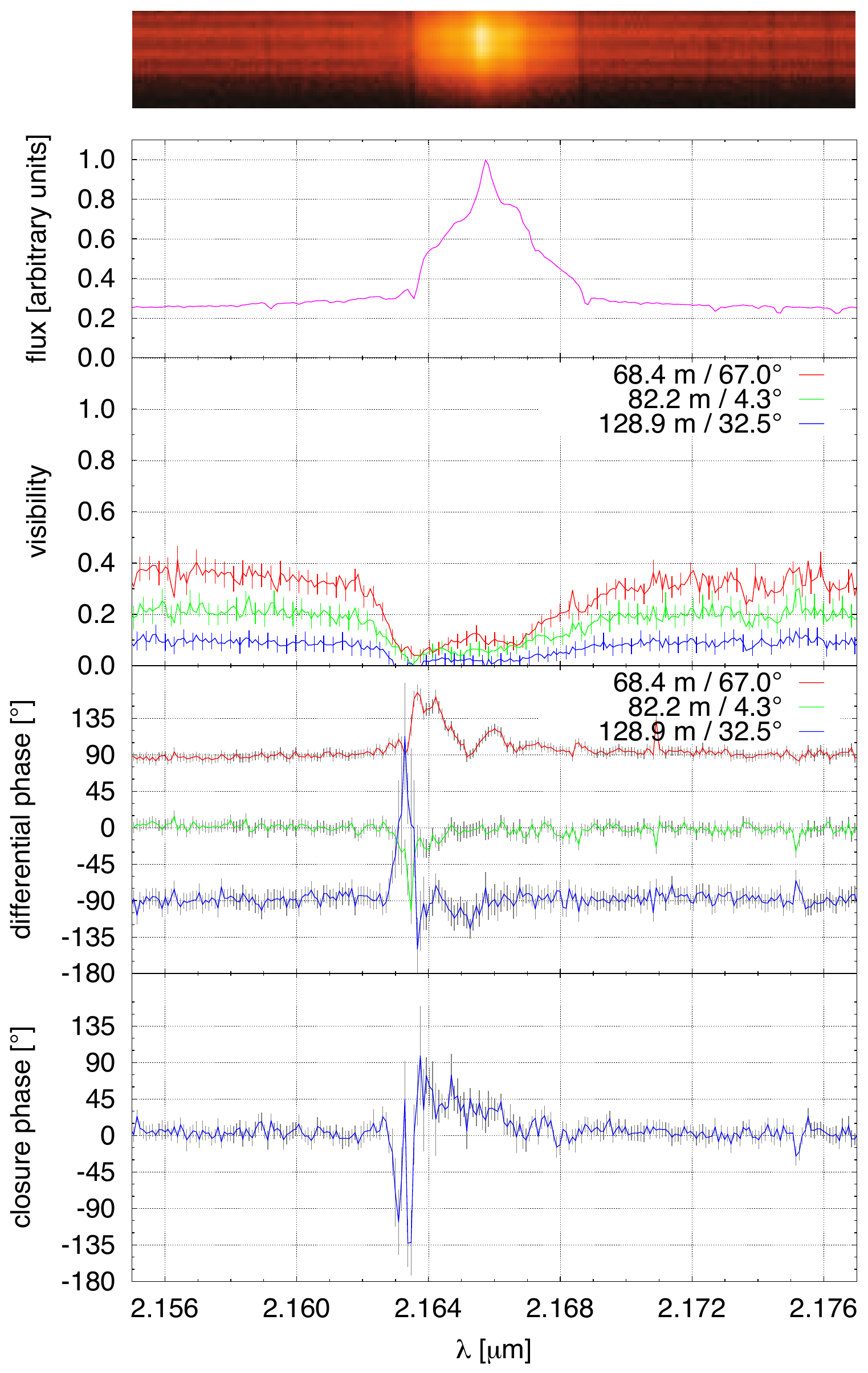}}  \caption{ Same as Fig.~\ref{obs1}, but different telescope configuration. }  \label{obs2} \end{figure}

In this paper, we present high spatial and high spectral resolution aperture-synthesis imaging of $\eta$~Car's innermost wind region.  We describe the observations in Section \ref{o}, while in  Section \ref{rr}, we present the velocity-resolved aperture-synthesis images reconstructed from the VLTI data. All images show the extended  wind region, which is more extended across the Br$\gamma$ line than in the continuum. In Section \ref{d}, we discuss  the results. Section \ref{s} is a  summary and conclusion.


\section{Observations and data processing}     \label{o}     

\subsection{Visibilities, wavelength-differential phases, and closure phases}  
    
The observations of $\eta$~Car listed in Table~\ref{list} were carried out with the ESO Very Large Telescope Interferometer  \citepads[VLTI;][]{2007NewAR..51..628S} and the AMBER interferometry instrument \citepads{2007A&A...464....1P}.  With projected baseline lengths up to about 129~m of the Auxilliary 1.8~m Telescopes (ATs) used, an angular resolution of $\sim$~6~mas ($\sim$~14~au)  was obtained in the $K$ band. Interferograms were recorded with the high spectral resolution mode (HR mode; spectral resolution $R = 12\,000$). Figure~\ref{uv} shows the $uv$ coverage of the observations.  The spatial-filter fibers in AMBER limit the FOV to the diameter of the fibers on the sky ($\sim$~250~mas for the ATs). The observations were carried out as part of the  OHANA  \citepads{2015IAUS..307..297R} backup target project (Observatory survey at High ANgular resolution of Active OB stars; Prog-ID: 092.D-0289). 

Figures~\ref{obs1} and \ref{obs2}  present two examples of high spectral resolution $\eta$~Car AMBER observations  (the dates of  observations shown in Figs.~\ref{obs1} and \ref{obs2} are 20 January 2014, 8:15 UT and 4 January 2014, 5:31 UT, respectively). The figures show, from top to bottom, an example of an AMBER interferogram, the line profile, the visibilities obtained at  three different baselines as a function of wavelength, the wavelength-differential phases at three baselines, and the wavelength dependence of the closure phase.  For the observations, the FINITO fringe tracker  \citepads{2008SPIE.7013E..18L}  was employed.  The observations were obtained between seven and five  months before the 2014.6 periastron passage (see Table~\ref{list}).  The observing  dates  correspond to an average  orbital phase of  $\sim$~0.90 if calculated as described in \citetads{2011ApJ...740...80M} and 0.91 if calculated as described in \citetads{2008MNRAS.384.1649D} or \citetads{2016ApJ...819..131T}.

We analyzed whether the interferometric  observables (spectrum,  visibilities, and phases of the object Fourier transform) changed over the two-month-long observing interval and found negligible changes. Therefore, all observations were combined to compute an averaged image. While the orbital eccentricity is $\sim$~0.90 \citepads{2001ApJ...547.1034C,2002A&A...383..636P}, the change in position of the companion relative to the primary star is small and the temporal evolution of the cavity is expected to be small over  the observing interval. The latter is substantiated by small changes in 3-D hydrodynamic models of the wind-wind cavity  that were computed at orbital phases 0.50, 0.90, and 0.952   \citepads{2012MNRAS.420.2064M, 2013MNRAS.436.3820M}. 

The data were reduced with our  own AMBER data processing software package, which uses the pixel-to-visibility matrix algorithm P2VM \citepads{2007A&A...464...29T,2009A&A...502..705C} in order to extract visibilities, differential phases, and closure phases for each spectral channel of an AMBER interferogram.  We derived 50 wavelength-dependent closure phases and 150 wavelength-differential phases from the interferograms. For only 14 of the observations are suitable calibrator star measurements available to calibrate the interferometric transfer function and the visibilities (see Table~\ref{list}). The number of calibrator measurements is small because the data were recorded within a backup-target programme (Prog-ID: 092.D-0289). The wavelength  calibration was performed using telluric lines. We used the same method as  described  in Appendix A of \citetads{2007A&A...464...87W}. 

The performance of the fringe tracker FINITO is usually different during target and calibrator observations, because the performance of FINITO depends on atmospheric conditions, target brightness, and target visibility.  This can lead to visibility calibrations of low quality. Therefore, we used the VLTI Reflective Memory Network Recorder (RMNrec) data  \citepads{2009A&A...493..747L,2012SPIE.8445E..1KM}  to improve the visibility calibration by rejecting 46\% of the  exposures that showed a disagreement between the RMNrec values of calibrator and science object. 

The calibrated continuum visibilities were fit with a two-dimensional exponential function in Fourier space, because the measured visibilities  and the Fourier transform of the Hillier continuum model curves are approximately exponential functions \citepads{2003A&A...410L..37V,2007A&A...464.1045K,2007A&A...464...87W}. From this visibility, we derived an axis ratio of 1.07 $\pm$ 0.14 and a PA  of the major axis of 159.5 $\pm$ 47$\degr$. The axis ratio and the PA are not well constrained  because the visibility errors are large and the number of calibrated visibilities is small. This model visibility function was used to calibrate the visibilities of the observations made without a calibrator (see Table~\ref{list}). 

\subsection{Differential-phase and closure-phase image reconstruction methods} 

\begin{figure*}   \centering    \includegraphics[width=160mm,angle=0]{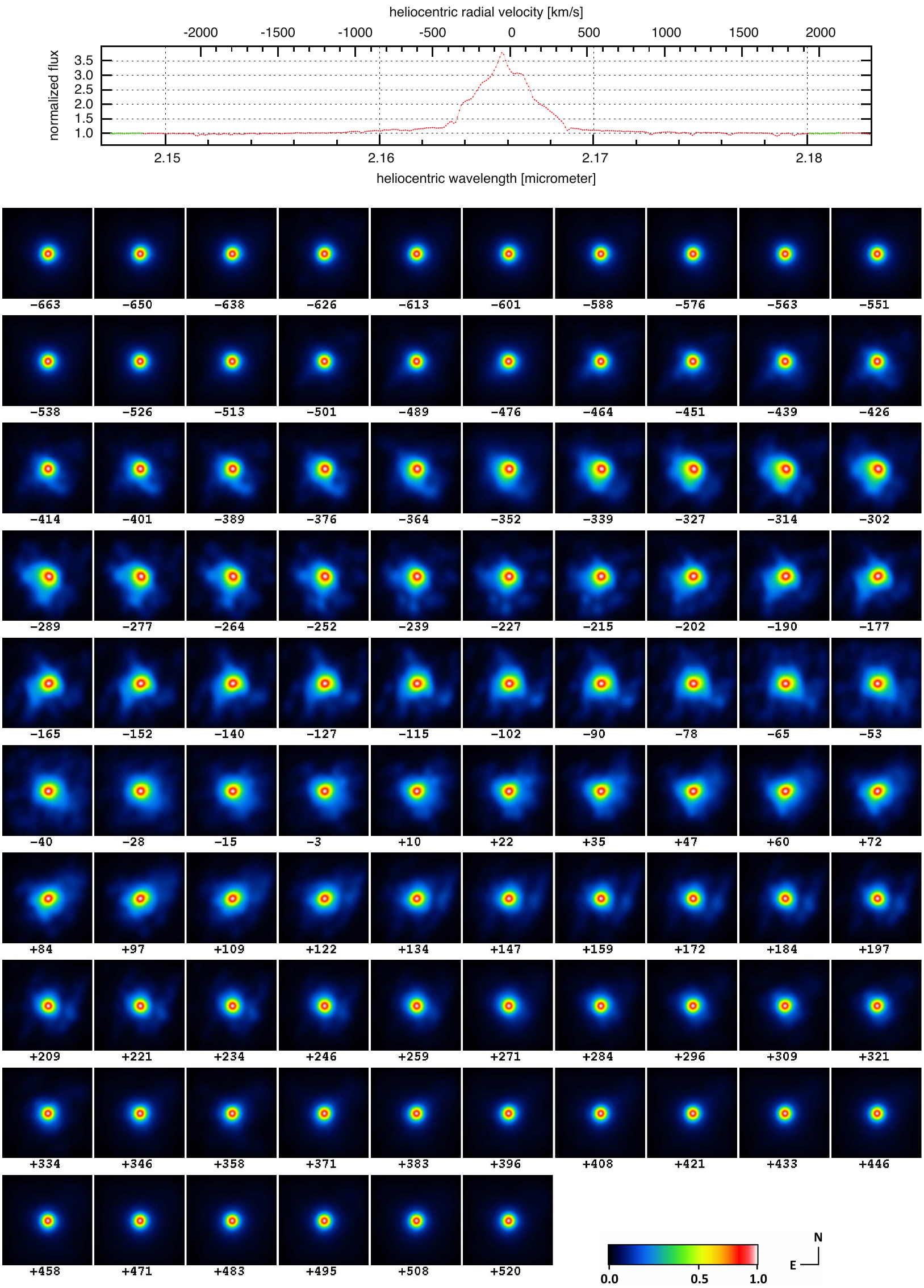}     
\caption{\small   Aperture-synthesis images of $\eta$~Car reconstructed with the differential-phase  method (see Sect.~\ref{o}; spectral resolution $R$ = 12\,000).  
{\it Top:} Continuum-normalized Br$\gamma$  line profile. The presented spectrum is the average of all spectra of the data listed in Table~\ref{list}. The small correction due to the system velocity of $-8~\kms$  \citepads{2004MNRAS.351L..15S} has been neglected.  {\it Bottom:}  Wavelength and velocity  dependence (velocities in units of $\kms$ below the images) of the reconstructed $\eta$~Car images across the Br$\gamma$ line. The FOV of the reconstructed images is 50 $\times$ 50~mas  (118 $\times$ 118~au; 1~mas corresponds to 2.35~au).  North is up, and east is to the left. In all images, the peak brightness is normalized to unity. At  radial velocities between approximately   $-$339 and $-252~\kms$, the images  are \emph{fan-shaped} and extended to the SE of the center of the continuum wind region.    }   \label{f} \end{figure*}

\begin{figure}   \centering    \includegraphics[width=82mm,angle=0]{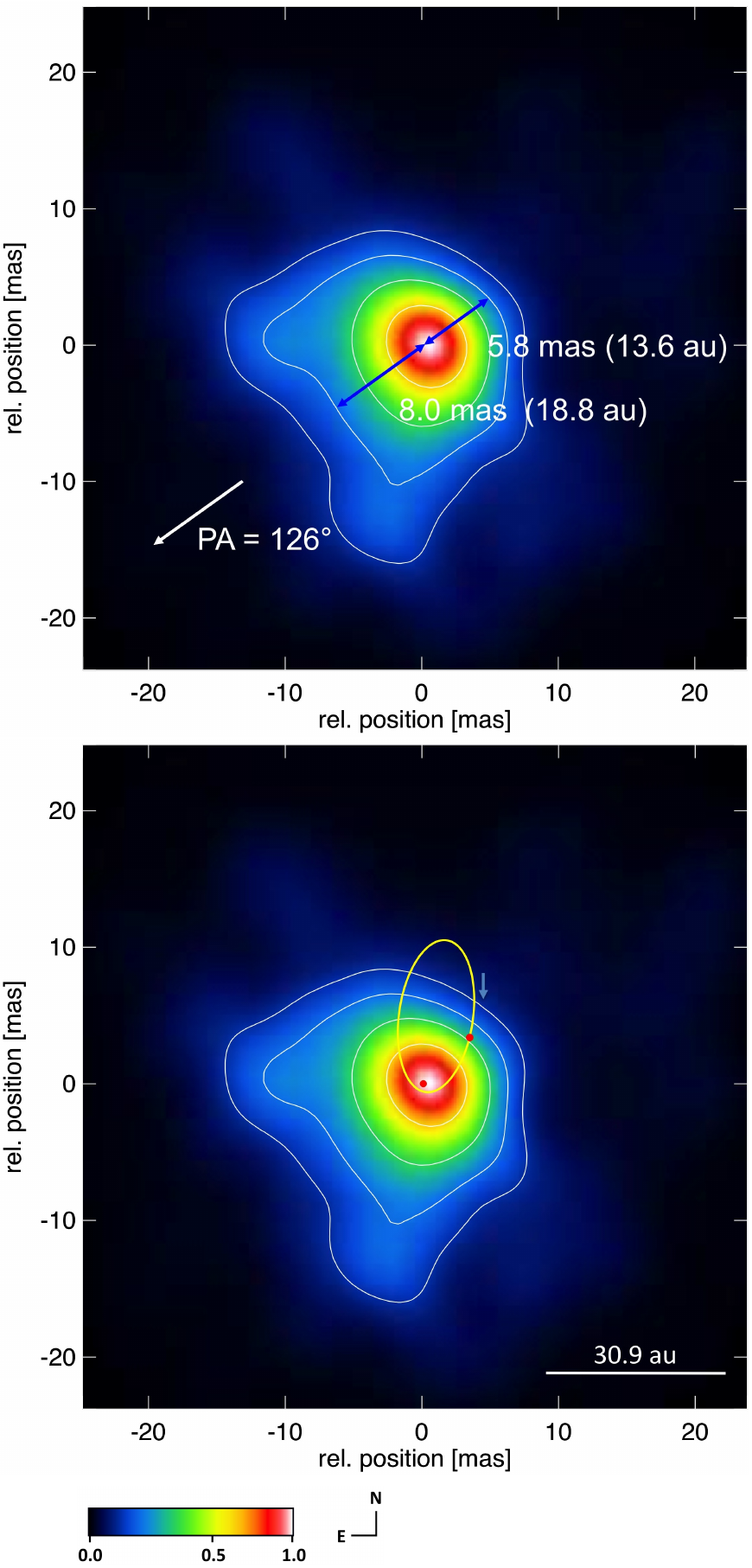} 
 
\caption[]{{\it Top:} Fan-shaped $\eta$~Car image at velocity of $-277~\kms$ (image from Fig.~\ref{f}).  The FOV of the  image is 50 $\times$ 50~mas  (118 $\times$ 118~au; the center of the continuum region is at coordinate zero).  Contour lines are plotted at  8, 16, 32, and 64\% of the peak intensity. The PA of the  symmetry axis of the fan is $\sim 126\degr$ (white arrow).  The fan-shaped structure extends $\sim 18.8$~au (8.0~mas; blue arrow) to the SE   and $\sim 13.6$~au (5.8~mas) to the NW,  measured with respect to the center of the continuum wind  along the fan symmetry axis at the 16\% intensity contour.   For comparison,  the radius $R_{\ast}$ of the primary star of $\eta$~Car is $\sim 100~R_{\sun}  \sim$ 0.47~au $\sim$ 0.20~mas  \citepads[$R_{\ast}$ is not well known; ][]{2001ApJ...553..837H}. Therefore, the huge wind extension to the SE   is about 40 times larger than $R_{\ast}$.      
{\it Bottom:} Overlay of a sketch of the orbit of the secondary star  relative to the primary star on the sky (yellow) and the fan-shaped  image at  $-277~\kms$. The orbit is adopted  from  \citeads{2016ApJ...819..131T} (phase 0.91 computed as described in this paper). The two red dots are the primary  and the secondary star at the time of our observations.  The binary separation and PA were approximately 4.9~mas (11.5~au) and  $315\degr$, respectively, at the time of observation (Sect.~\ref{d}).  The blue arrow indicates the motion direction and the bar at the bottom right  the length of the major axis of the orbit (30.9~au; \citeads{2012MNRAS.420.2064M}).      } \label{fan}   \end{figure}

We used two different image reconstruction methods,  called below  differential-phase method and    closure-phase method. If spectrally dispersed interferograms are available, the wavelength-differential phases in spectral channels across emission lines  can be used to derive   phases of the Fourier transform of the object  in spectral channels across  emission lines \citepads{
2007A&A...464....1P,      
2009ApJ...691..984S,     
2011A&A...526A.107M,  
2011A&A...529A.163O,  
2013A&A...555A..24O,   
2015A&A...577A..51M}.  
 This is possible if the phase of the Fourier transform of the object in the continuum is known (for example, if the continuum object is  unresolvable) or can be derived from a  continuum image  reconstructed from continuum visibilities and continuum closure phases using a closure phase method.  The reconstruction of images from a set of visibilities and phases of the object Fourier transforms derived from differential phases is discussed in  Appendix~\ref{ir}. This method is  called differential-phase  method in the following sections.  To reconstruct images using the differential-phase method, we used the minimization algorithm ASA-CG   \citepads{hager2006}, as described in \citepads{2014A&A...565A..48H}. The images presented in Figs.~\ref{f}, \ref{fan},  \ref{comp}, \ref{sub}, and \ref{sub2} were reconstructed with this differential-phase method.
 
Alternatively, we can reconstruct images from visibilities and closure phases  \citepads{1958MNRAS.118..276J}.  This method is briefly  called closure-phase method  in the following sections. We used the  \textit{IRBis}  method (Infrared Bispectrum image reconstruction method) for image reconstruction    \citepads{2014A&A...565A..48H}. The images presented in Fig.~\ref{b} in Appendix~\ref{ir} were reconstructed with the   closure-phase method. Data processing is discussed in  Appendix~\ref{ir} in more detail.

 
\section{Velocity-resolved aperture-synthesis images of $\eta$~Car's stellar wind and wind-wind collision zone  across the Br$\gamma$ line}     \label{rr}  

We reconstructed images  with both  methods (differential- and closure-phase method) discussed in  Sect.~\ref{o}. Both  methods were used to illustrate  the similarities and differences  of the reconstructions.

Figure~\ref{f} presents the images reconstructed from the observations listed in Table~\ref{list} using the  differential-phase  method (Sect.~\ref{o}).  The images clearly show  a strong wavelength dependence across the Br$\gamma$ line.  One of the most remarkable features of the images is  the fan-shaped structure extending to the southeast (SE; from PA $\sim$~90$\degr$ to $\sim$~180$\degr$) at velocities between approximately $-$376 and $-140~\kms$.  Figure~\ref{fan} shows the image at $-277~\kms$ to illustrate the asymmetric, fan-shaped structure and its size. As we will discuss in the next section, the size, structure, and velocity dependence  of these images suggest that we have  obtained  the first direct images of the  walls of the wind-wind collision cavity. Another interesting and unexpected structure is the  bar-like structure that extends to the southwest (SW) in the images at velocities from $-$426 to  $-339~\kms$.

The images presented in Fig.~\ref{b}  in Appendix~\ref{ir} were reconstructed with the closure-phase method  (see  Sect.~\ref{o}).  For the reconstructions in Fig.~\ref{b}, we used the same data set  as for the reconstructions shown in Fig.~\ref{f}. 

\begin{figure*}    \centering   \includegraphics[width=165mm,angle=0]{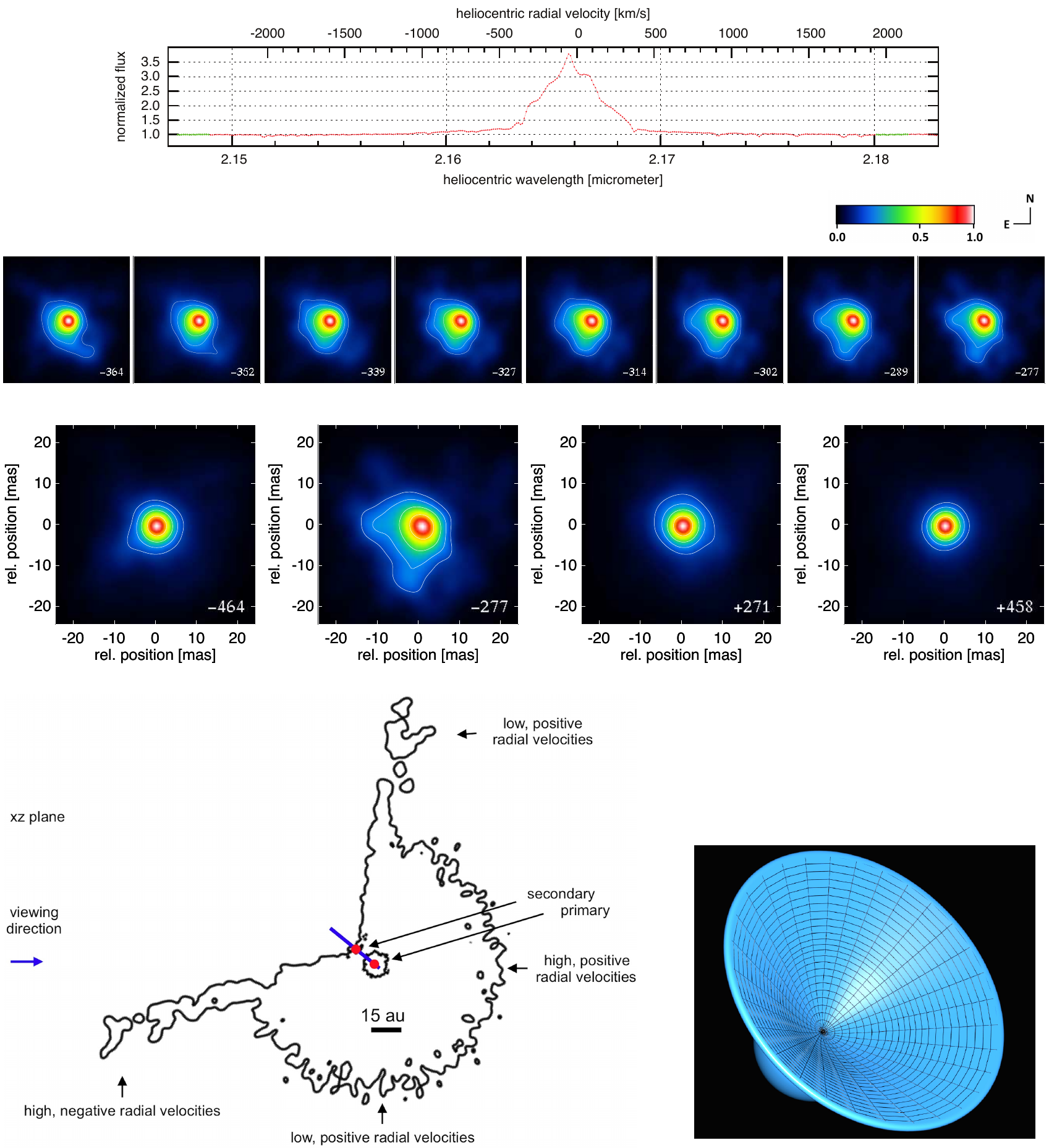}     \par\vspace{2mm}   \caption{\small Comparison of  $\eta$~Car  images  with model density distributions. 
{\it Top:} Continuum-normalized Br$\gamma$  line profile (same as in Fig.~\ref{f}). 
{\it Second row:} SW bar structure at  velocities of $-$364 to $-352~\kms$ and fan-shaped SE structure at velocities of about $-$339 to $-277~\kms$ (images from Fig.~\ref{f};  the FOV  is 50 $\times$ 50~mas  or 118 $\times$ 118~au; 1~mas corresponds to 2.35~au; north is up, and east is to the left). Contour lines are plotted at  8, 16, 32, and 64\% of the peak intensity. The fan-shaped SE structures   suggest that the reconstructed images are direct images of the wind-wind collision zone. 
{\it Third row:}  From left to right, images  at  velocities of $-464~\kms$ (continuum),   $-277~\kms$ (fan-shaped SE extension), +271~$\kms$, and +458~$\kms$ (continuum). At velocity of $-277~\kms$, the image is  asymmetric, fan-shaped, and more extended than the image at  velocity of +271~$\kms$. 
{\it Bottom left:} 
Illustration of the V-shaped wind-wind collision model, our viewing direction at the time of observation, and the radial velocities seen by the observer.  The iso-density contours are extracted from the model presented in Fig. B1  (phase 0.90) in  \citetads{2013MNRAS.436.3820M}.   Two density contours  (10$^{-16}$~g~cm$^{-3}$ and 10$^{-14}$~g~cm$^{-3}$) of the model density distribution are shown.   The binary orbit (semimajor axis length $\sim$~15.45~au) is sketched as a blue line. The projected elliptical orbit  is a line in the shown plane perpendicular to the orbital plane  \citepads{2012MNRAS.420.2064M,2013MNRAS.436.3820M}. The two red dots mark the positions of the primary and secondary star at the time of  observation (binary separation $\sim$~4.9~mas or 11.5~au; see Sect.~\ref{d}).
{\it Bottom right:}  
Line-of-sight view of a  three-dimensional  iso-density surface (density 10$^{-16}$~g~cm$^{-3}$) of the wind  at the time of observation (north is up, and east is to the left; FOV  is $\sim$50 $\times$ 50~mas).  This iso-density distribution is   a smoothed version of model iso-density derived by \citetads{2013MNRAS.436.3820M}. A small part of the Br$\gamma$ primary wind is visible to the SE. The continuum primary wind is not visible because its diameter is smaller than the Br$\gamma$ primary wind. The binary separation and PA were approximately 4.9~mas (11.5~au) and  $315\degr$, respectively, at the time of  observation. The secondary star is not visible in the images, because it is at least about 100 times fainter than the primary star, which is also not visible due to high extinction.          }\label{comp} \end{figure*}

Figures~\ref{f} to \ref{comp} present images that consist of both  continuum and Br$\gamma$ line components. To assist in the interpretation of the Br$\gamma$ line images, we additionally computed  the continuum-subtracted Br$\gamma$ line images shown in Figs.~\ref{sub} and \ref{sub2} in Appendix~\ref{cs} (these two figures show images of different velocity ranges). The white crosses are the centers of the continuum images. The images reconstructed with the differential-phase method (Fig.~\ref{f}) seem to have a slightly higher signal-to-noise ratio than the closure-phase images (Fig.~\ref{b}) because of the larger amount of Fourier phase information (i.e., 150 Fourier phases instead of 50 closure phases). Therefore,  the differential-phase  images were used to compute all  continuum-subtracted Br$\gamma$ line images. 

The continuum-subtracted images in Figs.~\ref{sub} and \ref{sub2} show the correct Br$\gamma$ brightness of the images, because they are not normalized to the peak brightness as the images in all previous figures. The computation of the continuum-subtracted images is discussed in Appendix~\ref{cs}. 

Surprisingly, in the continuum-subtracted images at highest negative  velocities ($\le -$563~$\kms$) in Fig.~\ref{sub}, the center of the Br$\gamma$ line emitting region is clearly shifted  about 1--2~mas to the  northwest (NW) of the center of the continuum image. To study this unexpected offset in more detail, we  computed the continuum-subtracted images shown in Fig.~\ref{sub2} (wider velocity range) in addition to the continuum-subtracted images in Fig.~\ref{sub} (notice the different color code required because of the faintness of most of the images).

The differential-phase (Fig.~\ref{f}) and closure-phase (Fig.~\ref{b}) images are similar, but there are  also differences. Both types of images present the discussed fan-shaped wind structure. To show the differences, we derived contour plot images of the differential-phase and closure-phase images at velocities between $-$376 and $-277~\kms$ shown in Fig.~\ref{co} in Appendix~\ref{com}.

Figure~\ref{phot1} in Appendix~\ref{wavelength} presents the velocity dependence of the photocenter shift of the images shown in Fig.~\ref{f}  (i.e., continuum plus line flux images) with respect to the center of the continuum images. Figure~\ref{phot2}  shows both the velocity  dependence of the size  of the images in Fig.~\ref{f} (measured as 50\% encircled-intensity radius) and the velocity dependence of the photocenter shift of the images  in Fig.~\ref{f}  with respect to the center of the continuum images. The biggest photocenter shifts are caused by the fan-shaped structure seen at many high negative velocities.

In Appendix~\ref{check},  we compare the measured visibilities, differential phases, and closure phases shown in Figs.~\ref{obs1} and \ref{obs2} with the visibilities and phases  derived from the images reconstructed with the differential-phase method to check whether the interferometric quantities of the reconstructed images approximately agree with the observations.


\section{Discussion}   \label{d}    
            
The reconstructed images in Fig.~\ref{f}  show a strong wavelength dependence of the intensity distribution across the Br$\gamma$ line. 
 \begin{itemize}   
 \item At wavelengths corresponding to radial velocities lower than about $-426~\kms$ and higher than about  +400~$\kms$, the continuum wind region is dominant.       
 \item At velocities between approximately $-$426 and  $-339~\kms$, a bar-like extension appears in the southwest (SW) of the continuum wind.      
 \item Between approximately $-$376 and $-140~\kms$, a  fan-shaped structure  is visible in the SE of the continuum wind.   This  unexpected fan-shaped structure is discussed in  Figs.~\ref{fan} and \ref{comp} in more detail.        
 \item  The wind intensity distribution is  more extended  at most high negative  velocities (e.g., $-$350 to $-250~\kms$) than at  high positive  velocities (e.g., +250 to +350~$\kms$).  
 \item At all  velocities, the continuum wind region is shining through the   Br$\gamma$ emitting wind region, suggesting that the line-emitting region is optically thin.
 \item The continuum-subtracted images in Figs.~\ref{sub} and \ref{sub2}  show that a faint  line-emitting region is visible at  high negative  and  positive velocities. At some high negative  velocities (e.g., about $-$563 to  $-825~\kms$ in Fig.~\ref{sub2}), the center of the line emission is offset to the  NW of the center of the continuum wind (see  Sect.~4.1).        \end{itemize}     
 
Figure~\ref{fan} illustrates the geometry of the fan-shaped  image at $-277~\kms$. Figure~\ref{comp} (second row) has the goal  to highlight the evolution of the fan-shaped structure between $-$364 and $-277~\kms$   and (third row) compare  images obtained at negative and positive velocities  to  illustrate the structure and size differences. For example, the  images  at  velocities of $-$277 and +271~$\kms$  clearly show a big size difference. This comparison and the cavity model sketch at the bottom of  Fig.~\ref{comp} are needed in the following discussions.

At  velocity  of $-277~\kms$,   the  reconstructed image has a fan-shaped intensity distribution (see Fig.~\ref{fan}).  The PA of the  symmetry axis of the fan is $\sim 126\degr$.   The fan-shaped structure extends $\sim$~8.0~mas (18.8~au) to the SE along a PA of $\sim126\degr$  and $\sim$~5.8~mas (13.6~au) to the NW along a PA of $\sim 306\degr$    measured   with respect to the center of the continuum wind image at the 16\% intensity contour.  

The observed image asymmetries indicate that {\it the LBV wind is not spherical or that there is an external influence on the wind}.  For the discussions in the next sections, we need to know the position of the secondary star on the sky at the time of  observation. The 3-D orientation of $\eta$~Car's orbit was derived by  \citetads{2012MNRAS.420.2064M}. Using these results and assuming the latest orientation parameters and orbit, as shown in Fig. 17 in \citetads{2016ApJ...819..131T}, we derived the projected binary separation of approximately 4.9~mas (11.5~au) and  PA of 315$\degr$ at the time of observation. The secondary star is not visible in our images, because it is at least about 100 times fainter than the primary star \citepads[e.g.,][]{2006ApJ...642..1098H,2007A&A...464...87W}. The primary is also not visible due to high extinction.

In the next sections, we discuss the high-velocity gas (Sect.~\ref{high}), the toroidal gas  (Sect.~\ref{tor}), the fossil wind (Sect.~\ref{fos}), the polar wind (Sect.~\ref{pol}), and the wind-wind collision (Sect.~\ref{wwcz}), because all of them may influence the observed images.

\subsection{High-velocity structure in the continuum-subtracted images}  \label{high}
In the continuum-subtracted images presented in Fig.~\ref{sub2}, the center of most of the Br$\gamma$ line images   is at the center of the continuum images (marked by white crosses). However, in the velocity range of about  $-$613 to  $-825~\kms$, the center of the Br$\gamma$ line emitting region is clearly shifted  about 1--2~mas to the  NW of the center of the continuum image. At $-663~\kms$, the NW offset is 1.7~$\pm$~0.4~mas along a PA of 292~$\pm$~20$\degr$. At velocities smaller than about $-837~\kms$, the high-velocity structure is still visible, but the NW offset disappeared. 

In the Br$\gamma$  line profile, the outer high-velocity region  is dominated by the electron scattering wings \citepads[e.g.,][]{2001ApJ...553..837H,2010A&A...517A...9G}. However, in the blue line wing, there may also be additional line contributions from  (1)  fast moving primary wind gas (in the wind cavity) accelerated by the fast secondary wind \citepads{2008MNRAS.386.2330D,2010A&A...517A...9G},  
and (2) the weak \ion{He}{i}~2.16127~$\mu$m line (the center of this line is approximately at velocity of about $-550~\kms$ in the Br$\gamma$ line profile). Model spectra of $\eta$~Car computed with the radiative transfer model of  \citetads{1998ApJ...496..407H} and \citetads{2001ApJ...553..837H} show  that the peak brightness of this \ion{He}{i}~2.16127~$\mu$m line is about 50 to 100 times weaker than the peak brightness of the Br$\gamma$ line.
 
The  cause of the observed $\sim$~1--2~mas (2.35--4.7~au)  NW offset of the line images at about  $-$613 to  $-825~\kms$  is unknown, but the high negative velocity and location of the high-velocity structure suggest that the emitting region (of both the above discussed high-velocity Br$\gamma$ emission and the \ion{He}{i}~2.16127~$\mu$m emission)  is located in the innermost region of the wind-wind collision cavity \citepads{2008MNRAS.386.2330D,2010A&A...517A...9G},  because  the innermost region of the  cavity was in the NW of the center of the continuum image at the phase of our observations   (binary separation and PA were 4.9~mas or 11.5~au and  315$\degr$ at the time of observation).

\subsection{Toroidal occulting gas and dust in the line-of-sight to the central object }   \label{tor}
Occulting gas in the line-of-sight to the central object may influence  the observed intensity distributions for the following reasons. The three ejected objects $\eta$~Car B, C, and D at separations between 0.1\arcsec  (235~au) and 0.3\arcsec\,  \citepads{1986A&A...163L...5W,1988A&A...203L..21H} were only about 12 times fainter than the central object in 1985 at the wavelength of about 900~nm. In the UV (at $\sim$~190~nm), where obscuring material leads to much more extinction, these objects B to D were only about three times fainter than the central object in 1991 \citepads{1995RMxAC...2...11W}. 

This seems to be puzzling because the  objects B, C, and D are objects illuminated by the central object (and only a very small  fraction of the star flux hits these objects). Therefore, several authors suggested that disk-like or toroidal occulting material   in our line-of-sight to the central object may partially obscure the central star, but not the objects B, C, and D, just like a natural coronagraph 
\citepads{1992A&A...262..153H,
1995RMxAC...2...11W,
1995AJ....109.1784D,
1997AJ....113..335D,
2001ApJ...553..837H,
2002ApJ...567L..77S,
2004ApJ...605..405S,
2010ApJ...710..729M,
2013MNRAS.436.3820M}. 

To study this occulting toroidal material, \citetads{1996A&A...306L..17F} performed speckle masking imaging polarimetry in the H$\alpha$ line. The  image obtained in polarized light  (see Fig. 2b in the polarimetry paper) shows a  bar with a length of $\sim$~0.4\arcsec \,(940~au) along a PA of $\sim 45\degr$ and $\sim 225\degr$ (about equal to the PA of the equatorial plane of the Homunculus).  It was interpreted as an obscuring equatorial disk. Interestingly, this polarized  bar has approximately the same length and PA as the fossil wind bar  discussed in the next section.  

Furthermore, in Fig.~\ref{f}, a bar- or arc-like structure  along PA of  $\sim 225\degr$  can  be seen at   velocities between approximately $-$426 and  $-339~\kms$  that may be caused by gas in the inner region of the equatorial disk or torus.

\subsection{Fossil wind}  \label{fos}
The  bar-like SW extension at PA of  $\sim 225\degr$ in Figs.~\ref{f} to \ref{comp} and  \ref{sub}  is mainly visible at  velocities between approximately $-$426 and  $-339~\kms$.  Its PA of $\sim 225\degr$   approximately agrees with the PA  of the  fossil wind structure reported by 
\citetads{2011ApJ...743L...3G,
2016arXiv160806193G} 
 and the polarisation bar \citepads{1996A&A...306L..17F}. The images of the  fossil wind  structure (at $\sim$~0.1--0.5\arcsec)  in \citetads{2011ApJ...743L...3G} were made by integrating the HST/STIS observations over the velocity range of $-$400 to  $-200~\kms$  (see also the  wind studies in \citeads{2009MNRAS.396.1308G, 2010RMxAC..38...52M,2013ApJ...773L..16T}). The fossil wind may slightly  influence the intensity distribution of the dominant inner wind region. However, the  fossil wind is faint compared to the innermost wind-wind collision zone.

Interestingly, both the intensity contours of the blueshifted \ion{Fe}{iii}~0.4659~$\mu$m  HST/STIS map  in \citetads{2010ApJ...710..729M} and the aforementioned blue-wing images in \citetads{2011ApJ...743L...3G}  show a similar asymmetric structure as some of our fan-shaped VLTI images, although of course HST's angular resolution was inferior.  \citetads{2010ApJ...710..729M}  discuss various origins of this structure, for example, the near side of a latitude-dependent structure in the outer stellar wind, the inner parts of the Homunculus ejecta, or the obscuring material that causes about two  magnitudes more the extinction in front of the primary star than in front of the ejecta at separations between 0.1 and 0.3\arcsec.  The fan-shaped images also look like inclined bipolar nebulae with a symmetry axis parallel to the Homunculus axis.

\subsection{High-density fast polar wind}  \label{pol}
 In the SE of the continuum center, we are looking at the south polar region of the stellar wind \citepads{2003ApJ...586..432S}
, because the inclination of the polar axis of the Homunculus nebula with the line-of-sight is $\sim 41\degr$  \citepads{
2001AJ....121.1569D, 
2006ApJ...644.1151S} 
 and the  PA  of the projected Homunculus axis is $\sim 132\degr$   \citepads{1997ARA&A..35....1D,
2001AJ....121.1569D,
2006ApJ...644.1151S}. 
\citetads{2003ApJ...586..432S} 
studied the stellar light reflected by different regions of the Homunculus nebula and found that the wind velocity and density of the massive primary wind of $\eta$~Car is higher at the polar region than at latitudes corresponding to our line-of-sight, and the polar wind axis is approximately parallel to the Homunculus axis. Therefore,  the SE extension of our images may at least partially  be caused by an increased wind density at the south pole.  On the other hand,  other studies suggest that the spectroscopic and interferometric observations can be explained by the wind-wind collision zone  \citepads{2012ApJ...759L...2G,2012ApJ...751...73M}.

\subsection{Wind-wind collision zone}  \label{wwcz}   

3-D smoothed particle hydrodynamic simulations of $\eta$~Car's colliding winds predict a bow shock with an enhanced density and an extended wind-wind collision cavity or bore hole  \citepads{2008MNRAS.388L..39O,
2009MNRAS.396.1308G,
2011ApJ...743L...3G,
2010A&A...517A...9G,
2010RMxAC..38...52M,
2012MNRAS.420.2064M,
2013MNRAS.436.3820M,
2015MNRAS.447.2445C,
2015MNRAS.450.1388C} 
that has a density distribution  \emph{more extended than the undisturbed primary wind region}, as in our observed images. This large extension of the wind-wind collision zone can, for example, be seen in the model density distributions presented in Fig. 18 (see insets) in \citetads{2009MNRAS.396.1308G}, 
Fig. 6 in \citetads{2012MNRAS.420.2064M}, 
Figs. B1 (orange region), 1, and 10 in \citetads{2013MNRAS.436.3820M}, 
Fig. 2 in   \citetads{2015MNRAS.447.2445C},   
and in the sketch in Fig.~\ref{comp}.                                                        

Because of the orbit orientation in space, the secondary star of $\eta$~Car  spends most of the time in front of the primary star  \citepads[e.g.,][]{2012MNRAS.420.2064M,2013MNRAS.436.3820M}.  Only for a few weeks around the periastron passage are the secondary star and the wind-wind collision zone behind the primary star (on the back or far side of the primary wind). According to aforementioned hydrodynamical simulations, at the time of  observation, the secondary star was in front of the primary star, and we were looking at the extended wind-wind collision cavity \citepads[e.g.,][]{2012MNRAS.420.2064M}. In other words, the wind collision cavity was  opened up into our line-of-sight at the time of  observation.  

Unfortunately, no $\eta$~Car model images of the Br$\gamma$ intensity distribution (e.g., 3-D smoothed particle hydrodynamic simulations combined with radiative transfer modeling) are available. Therefore, we use the density distribution of  3-D smoothed particle hydrodynamic simulations  (see above references) for the following discussion.  

Figure~\ref{comp} (bottom  left)  illustrates the wind-wind collision cavity and our viewing direction at orbital phase of 0.90. Two density contour  lines (density 10$^{-16}$~g~cm$^{-3}$ and 10$^{-14}$~g~cm$^{-3}$) of the wind collision cavity model  presented in Fig.  B1 in \citetads{2013MNRAS.436.3820M} are shown (bottom left).   The lower arm of the wind-wind collision cavity depicted in the sketch in Fig.\,\ref{comp}  (bottom left) can be seen in the reconstructed images (Figs.\,\ref{f} and \ref{comp}) to the SE of the center of the continuum wind, given the orientation of the orbit \citepads{2012MNRAS.420.2064M}. 

To illustrate the complicated 3-D wind structure, Fig.~\ref{comp} (bottom right) shows a  line-of-sight view of a smoothed three-dimensional  iso-density surface (density 10$^{-16}$~g~cm$^{-3}$)  of the wind  model  \citepads{2013MNRAS.436.3820M} at the time of  observation. In the next paragraphs, we will discuss the size and structure of the images at different velocities.

{\it (1) High  positive radial velocities.} At high positive  radial velocities, for example +200 to +400~$\kms$, the observed Br$\gamma$ wind intensity distributions are only slightly wider than the continuum wind region (see Figs.~\ref{f},  \ref{b}, \ref{sub}, and the  +271~$\kms$ image in the third row of Fig.~\ref{comp}).  The sketch at the bottom of Fig.~\ref{comp}  illustrates that the emission region corresponding to these high positive radial velocities is predicted to be on the back side of the primary wind (i.e., on the far side behind the primary star) and to have a {\it smaller angular diameter} than, for example, the emission region with very small  radial velocity (e.g., zero). The primary wind on the back side is not disturbed by the wind-wind collision zone at the orbital phase of our observations. We note that in all images in Figs.~\ref{f}  to \ref{comp} and \ref{b}, the peak brightness is normalized to unity. This explains why the compact wind at high positive  velocities  is not more visible in spite of the high line flux at these wavelengths.

{\it (2)  Low radial velocities.} The sketch in Fig.~\ref{comp}  shows that the  model region of small positive radial velocities (marked as "low, positive radial velocities" at the top of the sketch) is  the upper  cavity region. This upper cavity region is in the NW of the continuum center because of the orientation of the orbit \citepads{2012MNRAS.420.2064M}. This NW extension is only weakly visible in the reconstructed images in Fig.~\ref{f} (e.g., at velocities of +22 to +77~$\kms$). The reason may be the small cross-section of the outer region in our viewing direction (i.e., we are looking through only a small amount of material).  However, the faint low-velocity NW extension of the  intensity distribution of the wind collision zone is clearly visible in the continuum-subtracted images (see Fig.\ref{sub}). 

{\it (3)  High negative radial velocities in the range of approximately  $-$140 to $-376~\kms$.}  The sketch in Fig.~\ref{comp}  suggests that the wind collision cavity was opened up into our line-of-sight at the time of  observation, and that the emission from the lower arm of the cavity has  {\it high negative radial velocities}. The lower arm is in the SE of the center of the continuum wind because of the orientation of the orbit   \citepads{2012MNRAS.420.2064M}. At radial velocities between $-$215 and $-376~\kms$, the reconstructed images  (Fig.~\ref{f}) clearly show a fan-shaped SE extension as predicted by the wind-wind collision  model (sketch in Fig.~\ref{comp}).  This suggests that  our images   are   {\it direct images of the inner  wind-wind collision cavity}. Our observations provide detailed velocity-dependent image structures that can be used to test 3-D hydrodynamical, radiative transfer models of massive interacting winds.

This wind-wind cavity interpretation of our  images at negative velocities appears to be  more likely than the  high-density polar wind interpretation because a high-density polar wind would be expected to show a blue-shifted extension to the SE (near the south pole) and potentially a red-shifted extension to the NW (north pole). Instead we see a blue-shifted extension to the SE (at approximately $-$140 and $-376~\kms$) and no red-shifted extension to the NW at similar positive velocities (see Fig.~\ref{f}).  However, we cannot exclude that the additional emission sources discussed in Sect. 4.1 (e.g., the faint \ion{He}{i}~2.16127~$\mu$m emission line),  the fossil wind, toroidal occulting material in line-of-sight, and a high-density polar wind slightly influence  the observed wind cavity  intensity distributions.


\section{Summary and conclusions}    \label{s}     

We have presented the first VLTI-AMBER aperture-synthesis images of $\eta$~Car.  The angular resolution of the images is $\sim6$~mas ($\sim 14$~au).  Velocity-resolved images were obtained  in more than 100 different spectral channels distributed across the Br$\gamma~$2.166~$\mu$m  emission  line.   The images show that $\eta$~Car's stellar wind region is strongly wavelength-dependent and asymmetric. 

 At high negative velocities of $-$675 to about $-825~\kms$, the center of the Br$\gamma$ line emission is  located in the  NW of the center of the continuum image (offset $\sim$~1--2~mas or 2.35--4.7~au; Fig.~\ref{sub2}).  The  cause of this offset is unknown, but the high negative velocity and location of this high-velocity structure suggest that the emitting region  is located in the innermost region of the wind collision cavity \citepads{2008MNRAS.386.2330D,2010A&A...517A...9G},  because  both  the secondary star and the innermost region of the wind-wind collision cavity were located to the NW of the center of the continuum image at the phase of our observations. 
  
At  high positive radial velocities (e.g., +250 to +350~$\kms$), the reconstructed Br$\gamma$ images show a much narrower wind intensity distribution than at high negative radial velocities (e.g., $-$350 to $-250~\kms$), because at these positive velocities, we see the back side of the Br$\gamma$ primary wind behind the  primary star. This back-side  wind region is less extended than the front-side wind because it is less disturbed by the wind collision zone in front of the primary star.

The images  at velocities between approximately $-$140 and $-376~\kms$ show a large fan-shaped structure.  At  velocity  of $-277~\kms$ (see Fig.~\ref{fan}), the PA  of the  symmetry axis of the fan is $\sim126\degr$.   The fan-shaped structure extends $\sim 8.0$~mas (18.8~au) to the SE   and  $\sim 5.8$~mas (13.6~au) to the NW,     measured      along the fan symmetry axis at the 16\% intensity contour.  
   
3-D smoothed particle hydrodynamic simulations of $\eta$~Car's colliding winds  predict a large density distribution of the wind-wind collision zone that is more extended than the undisturbed primary wind.  At the time of our observations, the secondary star of $\eta$~Car  was in front of the primary star, and the cavity was opened up into our line-of-sight. The  observed  SE fan-shaped structure at  negative radial velocities (Figs.~\ref{f} to \ref{comp} and \ref{sub}) suggests that our  reconstructions  are {\it direct images of the  wind-wind collision zone}.  However,  the fossil wind, toroidal  occulting material in the line-of-sight, the faint \ion{He}{i}~2.16127$\mu$m emission line, and a high-density polar wind may also slightly influence  the observed image intensity distributions. Our observations provide detailed velocity-dependent image structures that can be used to test three-dimensional hydrodynamical, radiative transfer models of massive interacting winds.

Future  AMBER, GRAVITY, and MATISSE observations with a better uv coverage combined with 3-D smoothed particle hydrodynamic simulations will provide a unique opportunity to study the wind-wind collision zone of $\eta$~Car at many different lines in great detail.

\begin{acknowledgements} We thank all ESO colleagues for the excellent collaboration. The telluric spectra used in this work for spectral calibration of the AMBER data were created from data that was kindly made available by the NSO/Kitt Peak Observatory.  This publication makes use of the SIMBAD database operated at CDS, Strasbourg, France.   We thank the referee for helpful suggestions. AFJM is grateful for financial aid from NSERC (Canada) and FQRNT (Quebec).  S.K. acknowledges support from an STFC Rutherford Fellowship (ST/J004030/1) and ERC Starting Grant (Grant Agreement No. 639889). We thank Alexander Kreplin for helpful discussions.
 \end{acknowledgements}

 \bibliographystyle{aa} 
 \bibliography{ref} 

 
  \begin{appendix}

 
\section{Image reconstruction}    \label{ir}  

The goal of most image reconstruction methods is to find the  image that best agrees with the interferometric data measured at a set of $uv$ points (sparse $uv$ coverage). To reconstruct images using the differential-phase method and the closure-phase method discussed in Sect.~\ref{o}, we used, in both cases, the minimization algorithm ASA-CG   \citepads{hager2006}, as described in \citetads{2014A&A...565A..48H} . 

In {\it closure-phase methods}, the  interferometric input data are the measured bispectrum elements of the target, which are  calculated from the closure phases and visibilities measured at a set of $uv$  points. In {\it Fourier-phase methods}, as the differential-phase  method discussed in Sect.~\ref{o}, the interferometric data are a set of  measured Fourier spectrum elements calculated  from the differential phases and visibilities. 

To find the best target image, the $\chi^2$ function between the measured interferometric data and the corresponding data of the actual iterated image has to be minimized  in an iterative process. The $\chi^2$ function of  Fourier spectrum data is given by

\begin{equation} \chi^2 := \int \limits_{{\bf f} \in {\rm M^{(1)}}} \left|\frac{O_k({\bf f}) - O({\bf f})}{\sigma({\bf f})}\right|^2\, d{\bf f}, \label{equ:equ2}  \end{equation}

\noindent where $\rm M^{(1)}$ is the set of the two-dimensional spatial frequency coordinates ${\bf f}$ of all measured Fourier spectrum elements $O({\bf f})$ with their errors $\sigma({\bf f})$, and $O_k({\bf f})$ is the Fourier spectrum of the actual iterated image $o_k({\bf x})$ \citepads[e.g., Eq. 11 in ][]{1993A&A...278..328H}.

The $\chi^2$ function of the four-dimensional bispectrum data of the closure-phase method discussed in Sect.~\ref{o} is given by

\begin{equation}
\chi^2 := \int \limits_{{\bf f_1},{\bf f_2} \in {\rm M^{(2)}}} \left|\frac{O_k^{(3)}({\bf f_1},{\bf f_2}) - O^{(3)}({\bf f_1},{\bf f_2})}{\sigma({\bf f_1},{\bf f_2})}\right|^2\, d{\bf f_1}\,d{\bf f_2}, \label{equ:equ1}  \end{equation}

\noindent as described in  \citetads{2014A&A...565A..48H}. $\rm M^{(2)}$ denotes the set consisting of the four-dimensional spatial frequency coordinates $({\bf f_1},{\bf f_2})$ of all measured bispectrum elements $O^{(3)}({\bf f_1},{\bf f_2})$
with their errors $\sigma({\bf f_1},{\bf f_2})$, and $O_k^{(3)}({\bf f_1},{\bf f_2})$ is the bispectrum of the actual iterated image $o_k({\bf x})$ of the target.

Because of the sparse $uv$ coverage and the noise in the data, a weighted regularization function has to be added to the above  $\chi^2$ function (in both methods) and the resulting function, the cost function, has to be minimized, as described in  \citetads{2014A&A...565A..48H}.  

The IRBis processing of the $\eta$~Car data was performed as described in \citetads{2014A&A...565A..48H}. As start image of the reconstruction process, we used a  Lorentz function. The width of the Lorentz function in each spectral channel was determined by a fit to the measured visibilities. The FOV of the reconstruction region is 50~mas and the pixel grid used is 64 $\times$ 64 pixels. 

 The main image reconstruction parameters of IRBis are the size of a binary circular mask in image space and the strength of the regularization parameter. In order to find a good reconstruction, these two parameters are varied. The regularization functions "maximum entropy", "pixel difference quadratic", and "edge preserving" were tested. For the final reconstructions presented in this paper, we used the regularization function "pixel difference quadratic", which enforces smoothness in the reconstruction. For each regularization function, 18 different sizes of the binary circular mask and 30 different regularization parameters were tested. From all these reconstructions in each spectral channel, the reconstruction with the smallest value of the image quality parameter q$_{\rm{rec}}$ was chosen as image reconstruction result (q$_{\rm{rec}}$ is a quality measure including the $\chi^2$ values,  as described in  \citeads{2014A&A...565A..48H}).  The obtained images were convolved with the point spread function of a single-dish telescope with a diameter of 80~m.
 

\begin{figure*}    \centering   \includegraphics[width=160mm,angle=0]{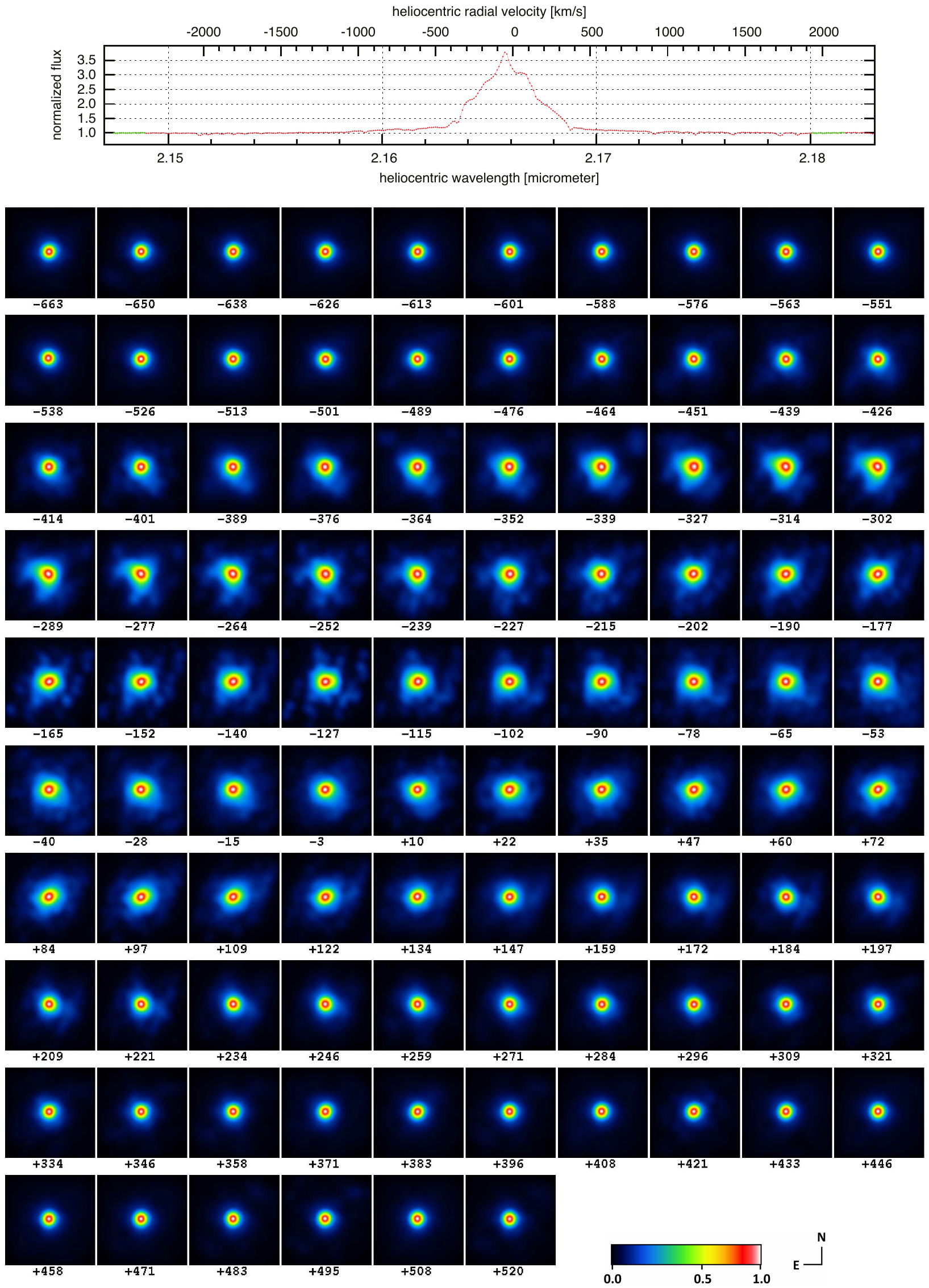}   \caption[]{\small Same as Fig.\,\ref{f}, but the  images were reconstructed with the closure-phase method (see Sect.\,\ref{o};  the FOV of the  images is 50 $\times $50~mas or 118 $\times $118~au; 1~mas corresponds to 2.35~au).  } \label{b}   \end{figure*}


\section{Continuum-subtracted Br$\gamma$ images}    \label{cs}  

\begin{figure*}  \centering    \includegraphics[width=160mm,angle=0]{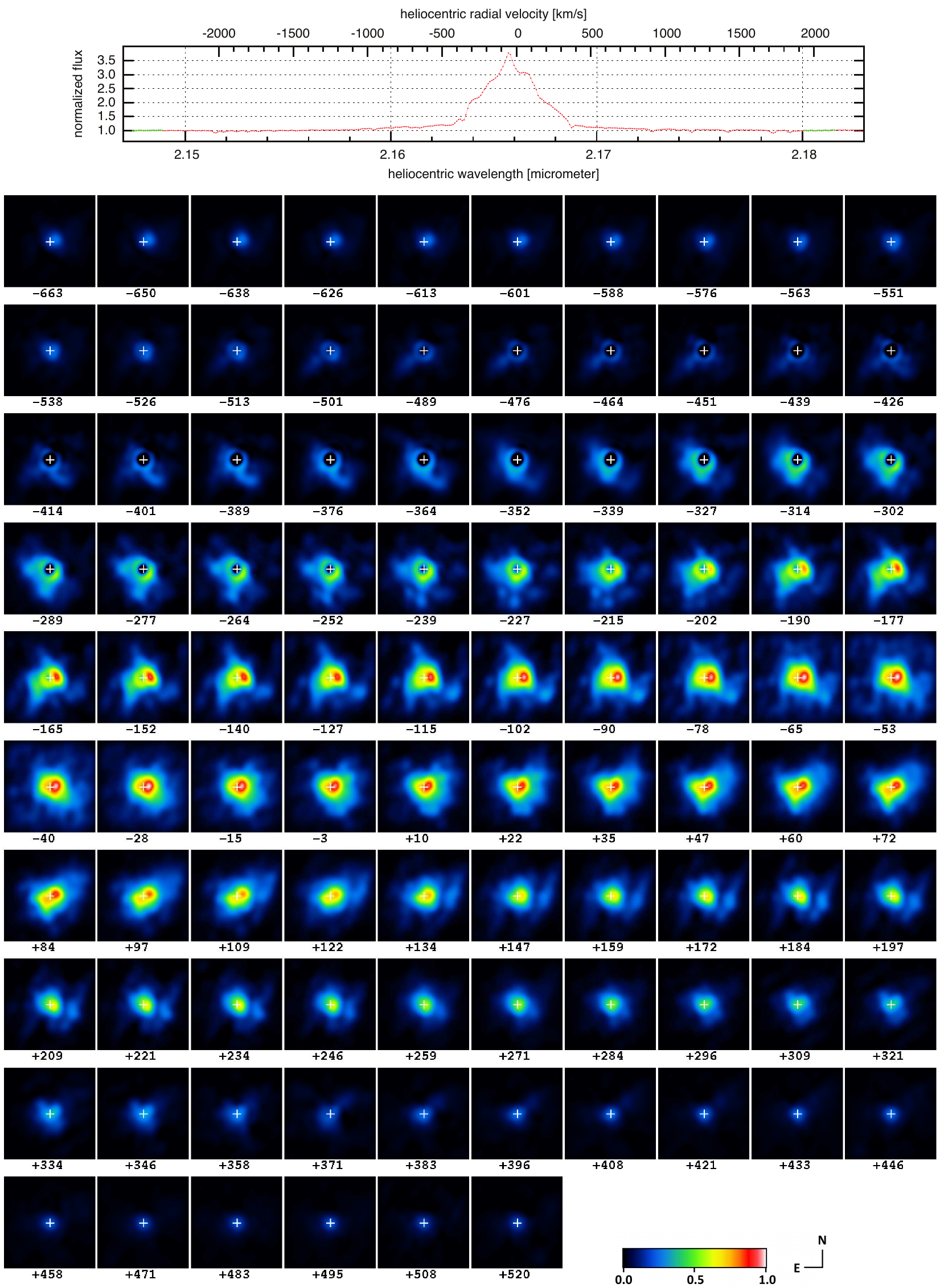}    \caption{Continuum-subtracted Br$\gamma$ images (see Sect.~\ref{d} and Appendix\,\ref{cs}; the FOV of the images  is 50 $\times$ 50~mas or 118 $\times$ 118~au).  The continuum-normalized spectrum at the top is the average of all spectra of the data listed in Table~\ref{list}. The white crosses mark the central position of the continuum wind. } \label{sub}  \end{figure*}

\begin{figure*}  \centering \includegraphics[width=146mm,angle=0]{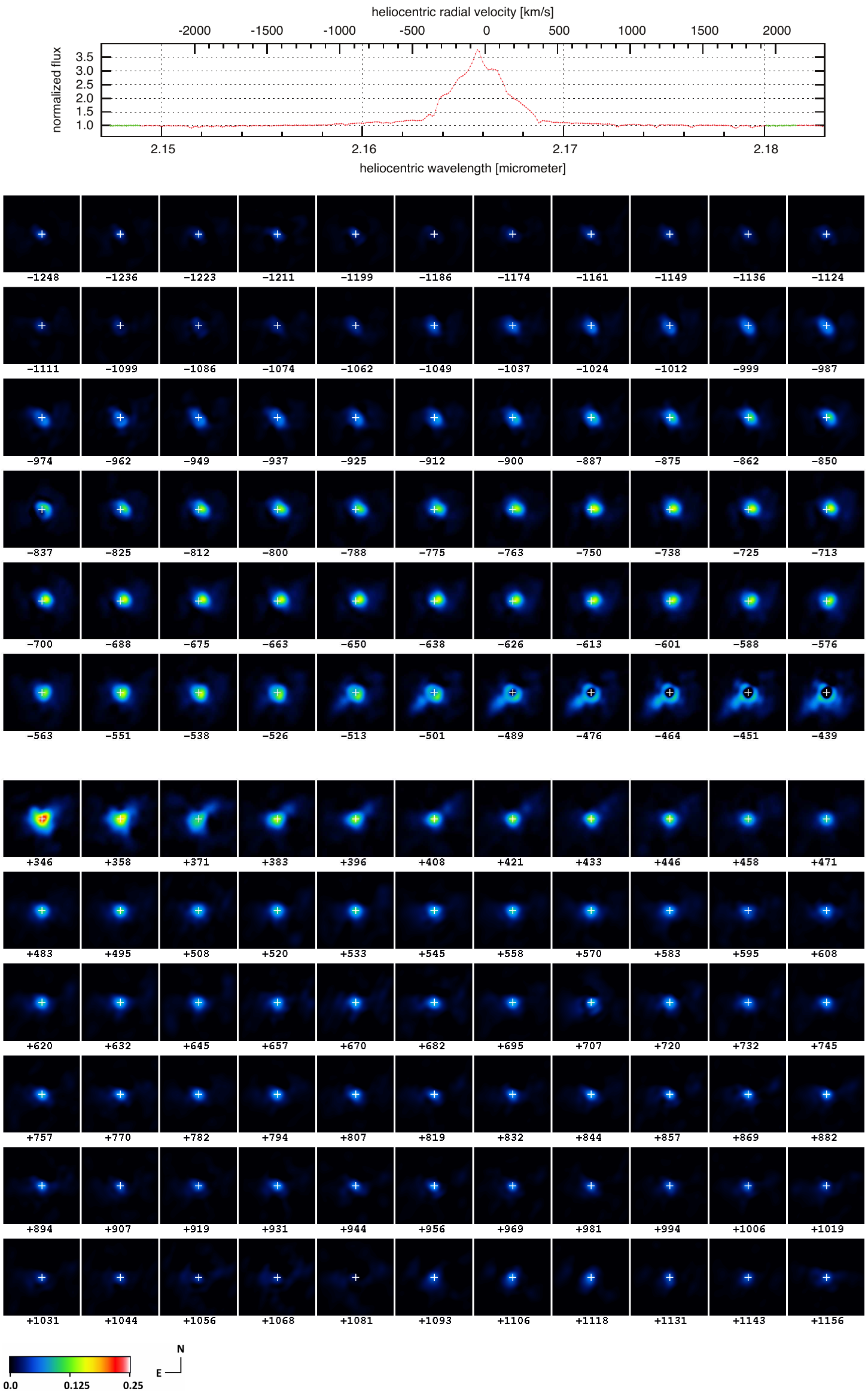}   \caption{Continuum-subtracted Br$\gamma$ images at  high negative and positive velocities (see Sect.~\ref{d} and Appendix\ref{cs}). A different color code is used in this figure compared to Fig.~\ref{sub} to enhance the faintest structures. The white crosses mark the central position of the continuum wind.} \label{sub2}  \end{figure*}

The image intensity distribution of the  images presented in Figs.~\ref{f} to  \ref{comp}, and \ref{b} consist of  the continuum and the line image components. In order to reveal the kinematics of signatures of the line-emitting gas more clearly, we additionally computed  continuum-subtracted images.  The continuum-subtracted  images are presented in Figs.~\ref{sub} and \ref{sub2}. They were computed in the following way. First, we computed the correct image  brightness (i.e.,  not normalized) by multiplying the differential-phase images in Fig.~\ref{f} with the flux of the line at the corresponding  velocity channel. From these images, we subtracted, in the next step, a continuum image (in differential-phase images,  the relative position between the continuum images and the line images is known). As continuum image, we have  computed  the average of the continuum images  at very high negative (in the range of  $-$2574 to $-2367~\kms$) and high positive  (+1954 to +2162~$\kms$)  velocities (i.e., outside the line Br$\gamma$; see green region of the spectrum in Fig.~\ref{sub2}).  The continuum flux at each wavelength channel was derived from a linear fit of the continuum flux values outside the line. 

As expected, dark gaps appear in  images at negative  velocities in the P Cygni absorption region. Obviously, the  continuum at these negative velocities   is smaller than the continuum outside the line because of P Cygni absorption. Therefore, the continuum-subtracted images do not show  correct line intensity distributions in the inner continuum wind region (i.e., a region as large as the continuum wind) at radial velocities  in the P Cygni absorption region. 

However, outside these inner regions,  the continuum-subtracted images allow us to see the Br$\gamma$ line intensity distribution with high contrast. For example,  the NW extension of the  intensity distribution caused by the wind-wind collision zone discussed in Sect.~\ref{wwcz}  (paragraph on small radial velocities) is better visible in the continuum-subtracted images than in the other images. The low-velocity cavity component is best visible in the images at  velocities in the range of  +35 to +109~$\kms$.

In the images at velocities between about $-$78 and $-28~\kms$ in Fig.~\ref{sub}, there are more small dots visible in the entire FOV than at other velocities. These dots are caused by the emission region that produces the well-known narrow emission line component at about $-40~\kms$. This narrow emission line component  in the Br$\gamma$ line of $\eta$~Car is caused by the  ejected, slow-moving  objects  B, C, and D \citepads{1986A&A...163L...5W,1988A&A...203L..21H,1995AJ....109.1784D,
1997AJ....113..335D} 
at separations from 0.1 to 0.3\arcsec (see Sect.~\ref{d}). The ejected object B is partially located in the AMBER FOV and there may exist additional fainter structures of similar nature. This extended intensity distribution cannot be imaged with the VLTI because the VLTI baselines are too long to sample visibilities and phases of extended structures larger than about 50~mas. Therefore, this large extended intensity distribution  breaks into many small dots in the reconstructed images. 


\section{Comparison of differential-phase and closure-phase images}    \label{com}   

\begin{figure*}   \resizebox{\hsize}{!}{\includegraphics{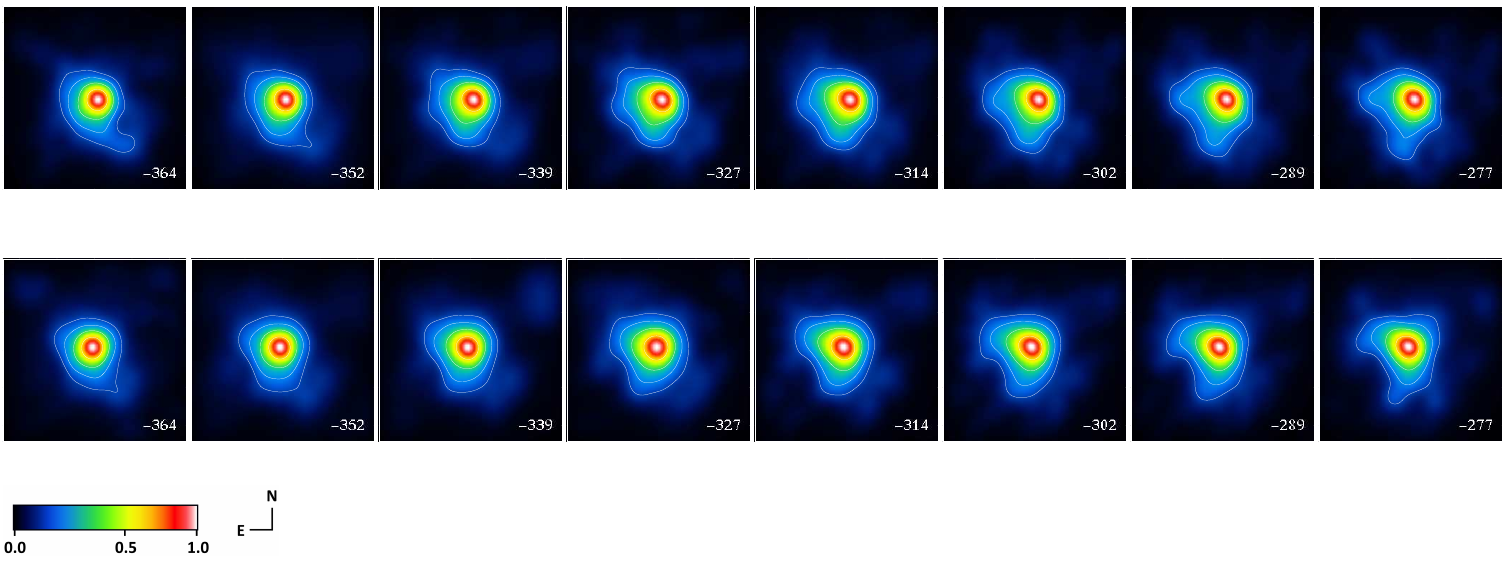}}   \caption{Comparison of the fan-shaped images reconstructed with the differential-phase (top) and closure-phase (bottom)  image reconstruction method   at velocities between $-$364 and $-277~\kms$   (the FOV of the images  is 50 $\times$ 50~mas or 118 $\times$ 118~au).  Contour lines are plotted at  8, 16, 32, and 64\% of the peak intensity.} \label{co}  \end{figure*}

In Fig.~\ref{co}, we compare contour line plots of the images reconstructed with the  differential-phase and closure-phase methods to illustrate that both types of images show the fan-shaped structure.


\section{Velocity dependence of size and photocenter shift}   \label{wavelength}
 
\begin{figure*}   \centering    \includegraphics[width=135mm,angle=0]{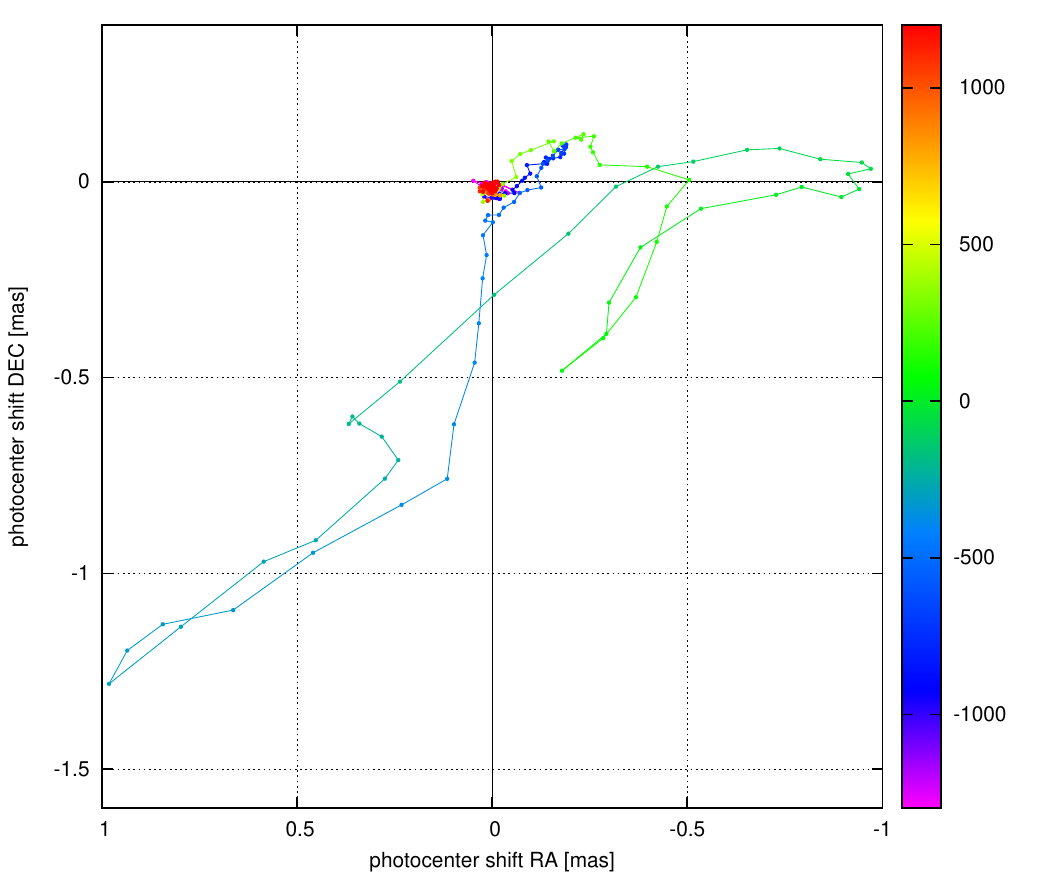} \\ 
\caption{Velocity dependence of the photocenter shift of the images shown in Fig.~\ref{f}  with respect to the center of the continuum images (velocities in the color bar are given in units of $\kms$.      } \label{phot1}  \end{figure*}

\begin{figure*}    \centering       \includegraphics[width=145mm,angle=0]{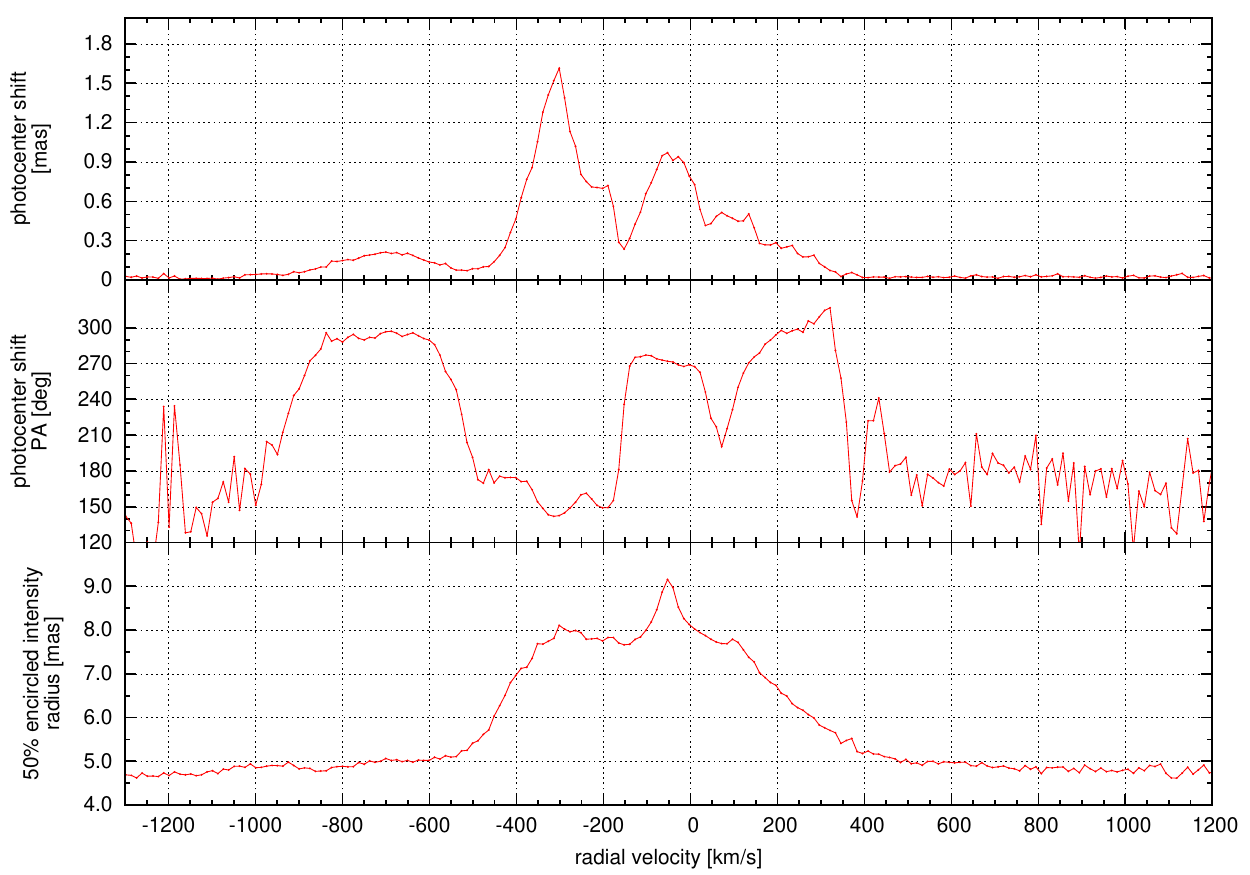} \\  
\caption{Velocity dependence of the size of the images  in Fig.~\ref{f}  (bottom) and   the photocenter shift (both amount and PA of the shift) of the images shown in Fig.~\ref{f}  with respect to the center of the continuum images.   }  \label{phot2}  \end{figure*}

Figure~\ref{phot1} presents the velocity  dependence of the  photocenter shift of the images  in Fig.~\ref{f} (i.e., images of continuum plus line flux) with respect to the center of the continuum images.  Figure~\ref{phot2} shows the velocity  dependence of both the image size and the  photocenter shift of the images in Fig.~\ref{f}  with respect to the center of the continuum images (velocities  in the color bar are given in units of $\kms$).  

\section{Comparison of observed interferometric quantities with the same quantities of the reconstructed images}   \label{check}
  
\begin{figure*}    \centering      \includegraphics[width=150mm,angle=0]{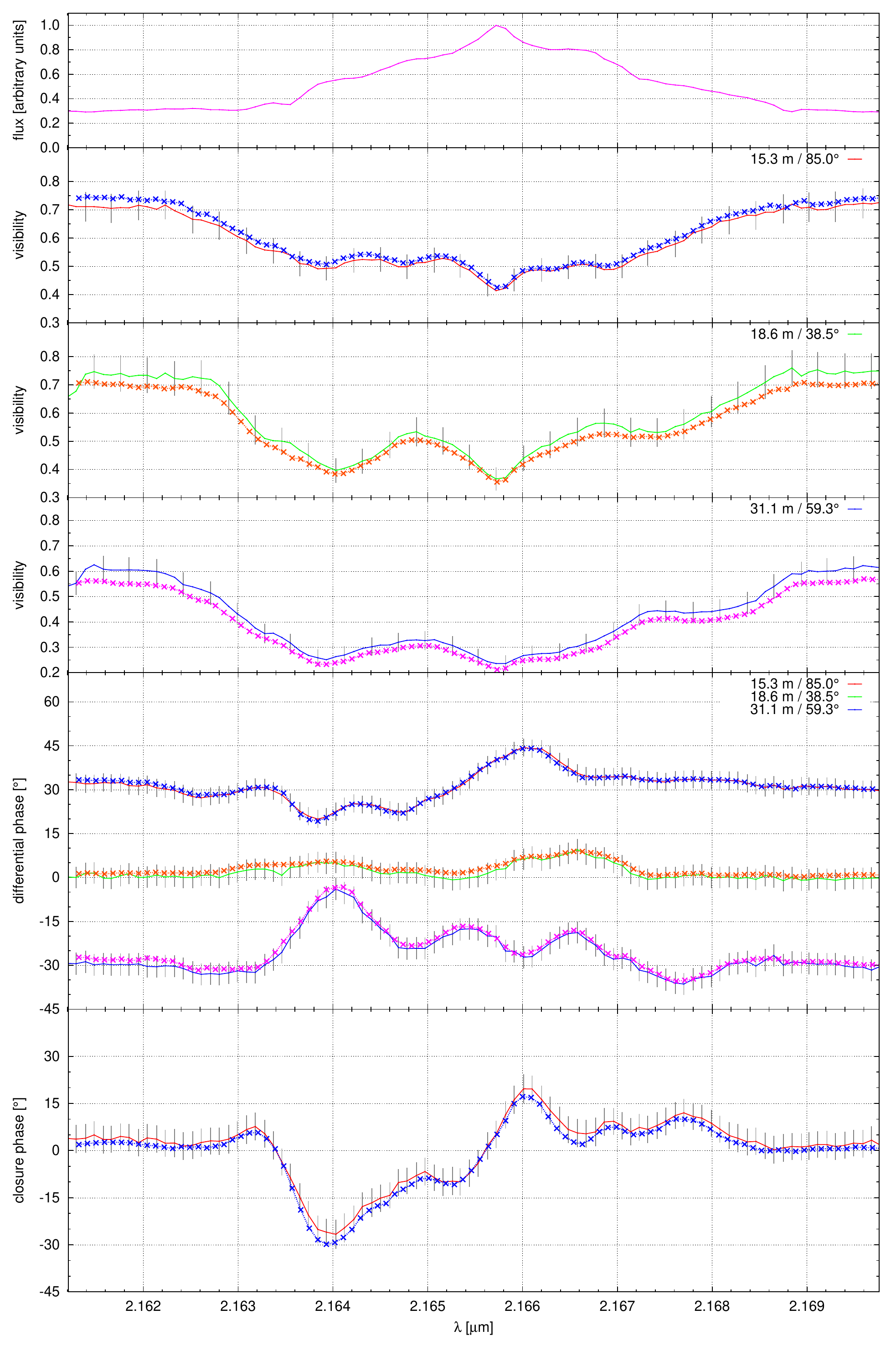} 
 \caption{Comparison of the  the measured visibilities, differential phases, and closure phases  shown in Fig.~\ref{obs1}  (curves with error bars) with the visibilities and phases  derived from the images in Fig.~\ref{f}   reconstructed with the differential-phase method (crosses). }    \label{q1} \end{figure*}

\begin{figure*}  \centering     \includegraphics[width=150mm,angle=0]{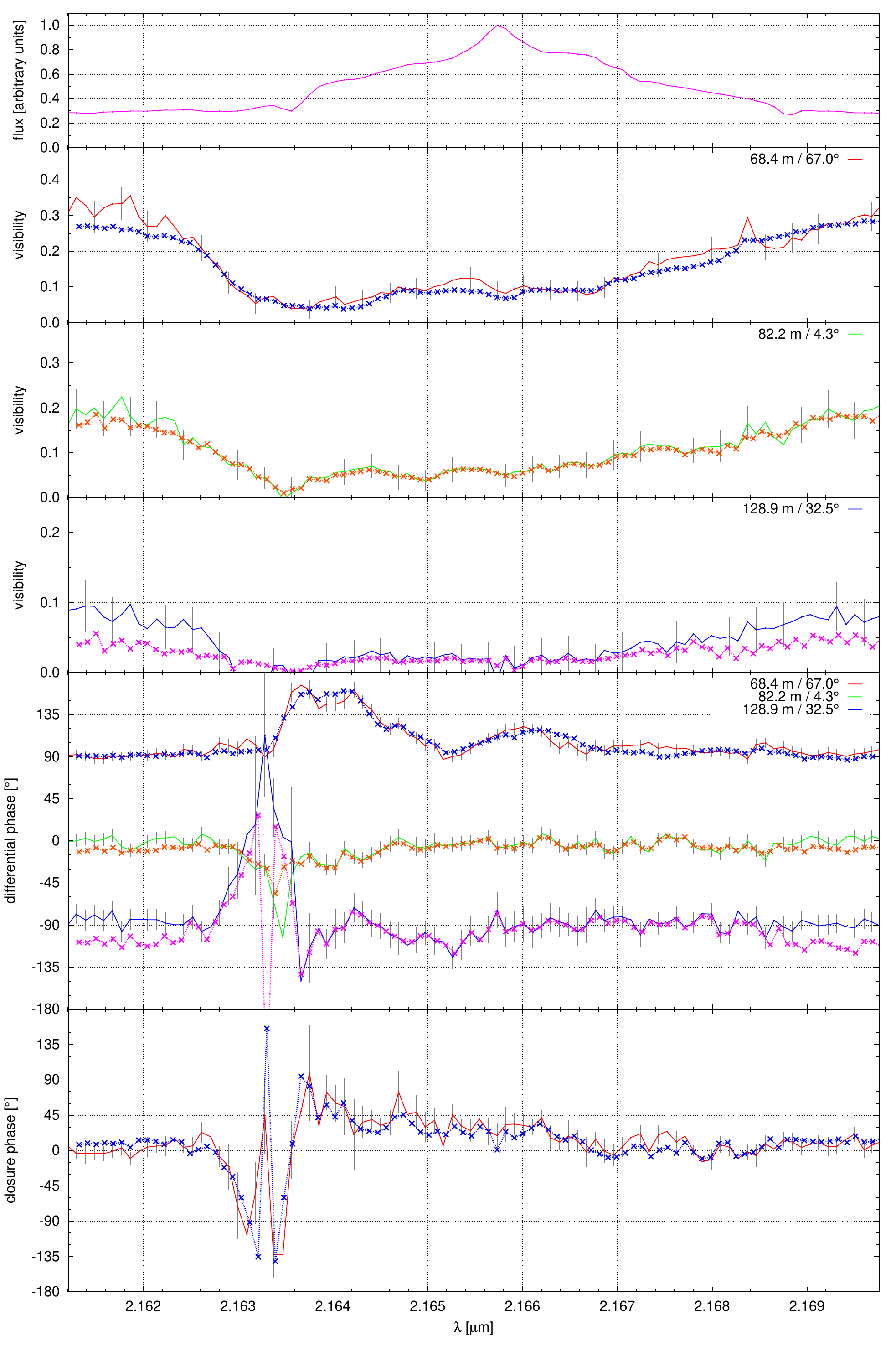}   
\caption{Same as Fig.~\ref{q1}, but comparison for Fig.~\ref{obs2}.      }  \label{q2}   \end{figure*}

In Figs.~\ref{q1}  and \ref{q2}, we compare the measured visibilities, differential phases, and closure phases shown in Figs.~\ref{obs1} and \ref{obs2}  with the visibilities, differential phases, and closure phases  derived from the images reconstructed with the differential-phase method to check whether the interferometric quantities of the reconstructed images approximately agree with the observations.

\end{appendix}

\end{document}